\documentclass[usenatbib]{mn2e}
\usepackage{epsfig}
\usepackage{times}
\def\gtrsim{~\rlap{$>$}{\lower 1.0ex\hbox{$\sim$}}}
\def\ltsim{~\rlap{$<$}{\lower 1.0ex\hbox{$\sim$}}}

\title[Dust and gas in NGC 2403]{The JCMT Nearby Galaxies Legacy
  Survey III: Comparisons of cold dust, polycyclic aromatic hydrocarbons, 
  molecular gas, and atomic gas in NGC 2403}
\author[G. J. Bendo et al.]
    {G. J. Bendo$^1$, C. D. Wilson$^2$, B. E. Warren$^{2, 3}$, 
    E. Brinks$^4$, H. M. Butner$^5$, P. Chanial$^{1,6}$,\newauthor 
    D. L. Clements$^1$, S. Courteau$^7$, J. Irwin$^7$, F. P. Israel$^8$, 
    J. H. Knapen$^{9, 10}$, J. Leech$^{11}$, \newauthor  
    H. E. Matthews$^{12}$, S. M\"uhle$^{13}$, G. Petitpas$^{14}$, 
    S. Serjeant$^{15}$, B. K. Tan$^{11}$, \newauthor
    R. P. J. Tilanus$^{16, 17}$, A. Usero$^{4,18}$, M. Vaccari$^{19}$, 
    P. van der Werf$^8$, C. Vlahakis$^8$, \newauthor
    T. Wiegert$^{20}$, M. Zhu$^{12,16}$\\
    $^1$Astrophysics Group, Imperial College, Blackett Laboratory,
        Prince Consort Road, London SW7 2AZ, United Kingdom\\
    $^2$Department of Physics \& Astronomy, McMaster University, Hamilton, 
        Ontario L8S 4M1, Canada\\
    $^3$ICRAR, M468, University of Western Australia, Crawley, WA, 6009, 
        Australia\\
    $^4$Centre for Astrophysics Research, University of Hertfordshire, 
        College Lane, Hatfield AL10 9AB, United Kingdom\\
    $^5$Department of Physics and Astronomy, James Madison University,
        911 Carrier Drive, Harrisonburg, VA 22807, USA\\
    $^6$Laboratoire AIM, CEA/DSM - CNRS - Université Paris Diderot, 
        Irfu/SAp, 91191 Gif-sur-Yvette, France\\
    $^7$Department of Physics, Engineering Physics and Astronomy,
        Queen's University, Kingston, Ontario K7L 3N6, Canada\\
    $^8$Sterrewacht Leiden, Leiden University, PO Box 9513, 2300 RA Leiden,
        The Netherlands\\
    $^9$Instituto de Astrof\'isica de Canarias, E-38200 La Laguna, Tenerife, 
        Spain\\
    $^{10}$Departamento de Astrof\'isica, Universidad de La Laguna,
        E-38205 La Laguna, Tenerife, Spain\\
    $^{11}$Department of Physics, University of Oxford, Keble Road, Oxford OX1
        3RH, United Kingdom\\
    $^{12}$National Research Council Canada, Herzberg Institute of Astrophysics,
        DRAO, P.O. Box 248, White Lake Road, Penticton, British Columbia\\
        V2A 69J, Canada\\
    $^{13}$Joint Institute for VLBI in Europe, Postbus 2, 7990 AA Dwingeloo, 
        The Netherlands\\
    $^{14}$Harvard-Smithsonian Center for Astrophysics, 60 Garden St., MS-78,
        Cambridge, MA 02138, USA\\
    $^{15}$Department of Physics \& Astronomy, The Open University, 
        Milton Keynes MK7 6AA, United Kingdom\\
    $^{16}$Joint Astronomy Centre, 660 N. A'ohoku Pl., University Park, Hilo, 
        HI 96720, USA\\
    $^{17}$Netherlands Organisation for Scientific Research, Laan van Nieuw 
        Oost-Indie 300, NL-2509 AC The Hague, The Netherlands\\
    $^{18}$Observatorio de Madrid, OAN, Alfonso XII, 3, E-28014 Madrid, Spain\\
    $^{19}$Dipartimento di Astronomia, Universit\'a di Padova, Vicolo 
        dell'Osservatorio 5, 35122 Padua, Italy\\
    $^{20}$Department of Physics and Astronomy, University of Manitoba, 
        Winnipeg, Manitoba R3T 2N2, Canada\\
}
\date{}
\pagerange{\pageref{firstpage}--\pageref{lastpage}}
\pubyear{}

\begin{document}
\label{firstpage}
\maketitle

\begin{abstract}
We used 3.6, 8.0, 70, 160~$\mu$m {\it Spitzer} Space Telescope data,
James Clerk Maxwell Telescope HARP-B CO~$J$=(3-2) data, National Radio
Astronomy Observatory 12 meter telescope CO~$J$=(1-0) data, and Very
Large Array H{\small I} data to investigate the relations among PAHs,
cold ($\sim20$~K) dust, molecular gas, and atomic gas within NGC~2403,
an SABcd galaxy at a distance of 3.13~Mpc.  The dust surface density
is mainly a function of the total (atomic and molecular) gas surface
density and galactocentric radius.  The gas-to-dust ratio
monotonically increases with radius, varying from $\sim100$ in the
nucleus to $\sim400$ at 5.5~kpc.  The slope of the gas-to-dust ratio
is close to that of the oxygen abundance, suggesting that metallicity
strongly affects the gas-to-dust ratio within this galaxy.  The
exponential scale length of the radial profile for the CO $J$=(3-2)
emission is statistically identical to the scale length for the
stellar continuum-subtracted 8~$\mu$m (PAH 8~$\mu$m) emission.
However, CO $J$=(3-2) and PAH 8~$\mu$m surface brightnesses appear
uncorrelated when examining sub-kpc sized regions.  
\end{abstract}


\section{Introduction}

The James Clerk Maxwell Telescope (JCMT) Nearby Galaxies Legacy Survey
(NGLS) is an 850~$\mu$m and CO~$J$=(3-2) survey with the goals of
studying molecular gas and dust in a representative sample of galaxies
within 25~Mpc.  The sample contains three subsets: a subset of 31
galaxies from the {\it Spitzer} Infrared Nearby Galaxies Survey
\citep[SINGS;][]{kabetal03} sample; an H{\small I} flux-limited sample
of 36 galaxies in the Virgo Cluster; and an H{\small I} flux-limited
sample of 72 field galaxies.  At this point in time, CO~$J$=(3-2)
observations of the SINGS galaxies and Virgo Cluster galaxies have
been completed.  Aside from the 3.6-160~$\mu$m data from SINGS that
resolve structures less than a kiloparsec in size in the closest
galaxies, H{\small I} Nearby Galaxies Survey
\citep[THINGS;][]{wbbbktl08} are also available for many of these
galaxies.  Together, these data are a powerful dataset for studying
the properties of the interstellar medium (ISM) on sub-kiloparsec
scales within nearby galaxies such as variations in gas-to-dust ratios on
sub-kiloparsec scales.

Gas-to-dust ratios within nearby galaxies have been a major topic of
research since the launch of the Infrared Astronomical Satellite
(IRAS).  Debate on this subject began when dust masses derived from
IRAS 60 and 100~$\mu$m data seemed a factor of 10 too low to match
what was expected either from the depletion of metals from the ISM or
from the ratio of gas column density to dust extinction
\citep[e.g.][]{yskl86, yxkr89, dy90}.  More recent calculations of
global gas and dust masses using {\it Spitzer} Space Telescope
\citep{wetal04} and James Clerk Maxwell Telescope data for spiral
galaxies produce gas-to-dust mass ratios of $\sim150$
\citep[e.g][]{rtbetal04, betal06, ddbetal07}.  Although the measured
ratios are reliant upon CO to H$_2$ mass conversion factors that do
not vary with metallicity or star formation activity, the ratios that
were measured are close to what is expected from the ratio for the Milky
Way derived from the comparison of gas column densities to optical
extinction \citep[e.g.][]{k03, w03} or from the depletion of metals
from the ISM \citep[e.g.][]{w03, l04, k08}.  Therefore, the natural
expectation would be that dust emission from kiloparsec-sized regions
should be correlated with the sum of atomic and molecular gas emission
on those spatial scales.  However, since the gas-to-dust ratio depends
on metallicity and since metallicity varies with radius within most
galaxies \citep[e.g.][]{ve92, zkh94, vetal98}, the gas-to-dust ratio
should decrease with radius.  For the most part, these assumptions are
only beginning to be tested, mainly because the combination of
H{\small I}, CO, and far-infrared dust emission data that is resolved
on kiloparsec scales has not been available for nearby galaxies.

Related to this topic is the association between polycyclic aromatic
hydrocarbon (PAH) emission in the mid-infrared and other constituents
of the cold ISM.  Several observational studies using data from either
the Infrared Space Observatory \citep[ISO; ][]{ketal96} or {\it
  Spitzer} have shown that, in nearby spiral galaxies, PAH emission is
strongly correlated with cold ($\sim20$ K) dust emission in the
far-infrared, at least on scales $\gtrsim1$~kpc \citep[e.g.][]{mtl99,
  hkb02, betal06, betal08, zwcl08}.  Additionally, \citep{retal06}
found that PAH emission is correlated with CO spectral line emission
at submillimetre and millimetre wavelengths within nearby spiral
galaxies, although these results are based on using radial averages.
These results strongly suggested that PAH emission is a tracer of both
dust and gas in the cold interstellar medium (ISM). 

Some of the SINGS/THINGS/JCMT NGLS galaxies have already been used in
comparisons of H{\small I} and CO emission with star formation,
including measures of star formation that incorporate 24~$\mu$m
emission \citep[e.g.][]{ketal07, blwbdmt08, wetal09}.  These studies
have found that star formation is correlated with molecular gas
surface density or with a sum of molecular and atomic gas surface
density; no correlation exists between star formation and H{\small I}
emission by itself.  Since 24~$\mu$m dust emission is included in
these comparisons, the results imply that dust emission should be
correlated with molecular or total gas surface densities but not with
atomic gas surface densities, although a direct comparison between
dust, atomic gas, and molecular gas is needed.

As a first look into these topics with the combined SINGS, THINGS, and
JCMT NGLS data, we compare PAH, cold dust, CO~$J$=(3-2), and H{\small
  I} on sub-kiloparsec scales for NGC~2403.  This SABcd spiral galaxy
\citep{ddcbpf91} was selected because it is relatively nearby
\citep[the distance to the galaxy is $3.13 \pm 0.14$~Mpc; ]{fetal01},
the galaxy and the dust emission is very extended \citep[the optical
  disc is 21.9~arcmin $\times$ 12.3~arcmin; ][]{ddcbpf91}. While the
inclination is $62.9^\circ$ \citep{dwbtok08}, the galaxy is close enough to
face-on that it is possible to distinguish spiral structures and
individual star-forming regions.  As is found in similar late-type
spiral galaxies, the regions with the strongest star formation are
located outside the centre of of the galaxy \citep[e.g][]{drms99,
  betal08}, while the cold dust emission peaks in the centre
\citep{betal08}.  Consequently, emission related to star formation can
be easily disentangled from emission related to dust surface density.

\section{Observations and data reduction}

\subsection{Mid-infrared data}

We use the 3.6 and 8.0~$\mu$m data taken with the Infrared Array
Camera \citep[IRAC;][]{fetal04} on {\it Spitzer} as part of SINGS. The
observations consisted of a series of 5~arcmin $\times$ 5~arcmin
individual frames taken in a mosaic pattern that covers a 25~arcmin
$\times$ 25~arcmin region.  The data used for this analysis come from
astronomical observation requests (AORs) 5505792 and 5505536, which
were separated by four days; these separate observations allow for the
identification and removal of transient phenomena, particularly
asteroids.  The full-width half-maxima (FWHM) of the point spread
functions (PSFs), as stated in the Spitzer Observer's Manual
\citep{sscmanual07}\footnote{http://ssc.spitzer.caltech.edu/documents/som/},
are 1.7 and 2.0~arcsec at 3.6 and 8.0~$\mu$m, respectively.  Details
on the observations can be found in the documentation for the SINGS
fifth data delivery \citep{sings07}\footnote{\raggedright Available at
http://data.spitzer.caltech.edu/popular/sings
/20070410\_enhanced\_v1/Documents/sings\_fifth\_delivery\_v2.pdf}.

The final images were created from version 14 of the basic calibrated
data frames produced by the Spitzer Science Center.  The data were
then processed through the SINGS IRAC pipeline, which applies
distortion and rotation corrections, adjusts the offsets among the
data frames, removes cosmic rays, subtracts the background, and then
combines them together using a drizzle technique.  A description of
the technique is presented in \citet{retal06} and \citep{sings07}.
The flux calibration is expected to be accurate to 3\%
\citep{sscmanual07}.

The final IRAC images include a residual background with a gradient.
We interpolated a smoothed version of the background outside the
optical disc of the galaxy to determine the background levels within
the optical disc, and we then subtracted this background from the
data.  We then masked out regions contaminated by emission from bright
foreground stars (identified as unresolved sources in the unconvolved
IRAC data with 3.6~$\mu$m/8~$\mu$m surface brightness ratios $\gtrsim
5$).  To correct for the diffusion of light through the IRAC detector
substrate, the 3.6 and 8.0~$\mu$m data are multiplied by the
``infinite'' aperture corrections described by \citet{retal05}, which
should not be confused with the types of aperture corrections used for
measuring the flux densities of unresolved sources.  The correction
factors are 0.944 and 0.737 at 3.6 and 8.0~$\mu$m, respectively.  We
then subtracted the stellar continuum from the 8~$\mu$m surface
brightness maps (in MJy sr$^{-1}$) using
\begin{equation}
I_\nu(PAH~8\mu m)=I_\nu(8\mu m)-0.232I_\nu(3.6\mu m),
\label{e_8starsub}
\end{equation}
which was derived by \citet{hraetal04}.  We will henceforth refer to this as
PAH 8~$\mu$m emission so as to distinguish it from the 8~$\mu$m emission 
that includes stellar continuum emission.

\subsection{Far-infrared data}
\label{s_data_dust}

We used the Multiband Imaging Photometer for Spitzer \citet[MIPS;
][]{ryeetal04} 70 and 160~$\mu$m data taken of NGC~2403 as part of
SINGS.  These observations are composed of the AORs 5549568 and
5549824.  These AORs are scan map observations of the galaxy that were
performed at the medium scan speed.  Each scan map is $\sim0.5^\circ$
wide and $1^\circ$ long, which is sufficiently wide to cover the
entire optical disc of the galaxy.  The two AORs were executed three
days apart so as to allow for the identification and removal of
transient sources, although most detectable sources usually affect
only the 24~$\mu$m band.  The FWHM of the PSFs are 18~arcsec at
70~$\mu$m and 38~arcsec at 160~$\mu$m \citep{sscmanual07}.  The
documentation for the SINGS fifth data delivery \citep{sings07}
contains additional details on the observing strategy.

The 70 and 160~$\mu$m images were created from raw data frames using
the MIPS Data Analysis Tools \citep[MIPS DAT;][]{getal05} version 3.10
along with additional processing steps.  First, ramps were fit to the
70 and 160~$\mu$m reads to derive slopes.  In this step, readout jumps
and cosmic ray hits were also removed, and an electronic nonlinearity
correction was applied.  Next, the stim flash frames (frames of data
in which a calibration light source was flashed at the detectors) were
used as responsivity corrections.  After this, the dark current was
subtracted from the data, and an illumination correction was applied.
Following this, short term variations in the signal (often
referred to as drift) were removed, and additional periodic variations
in the background related to the stim flash cycle were subtracted;
this also subtracted the background from the data.  Next, a robust
statistical analysis was applied to cospatial pixels from different
frames in which statistical outliers (which could be pixels affected
by cosmic rays) were masked out.  Once this was done, final mosaics
were made using pixel sizes of 4.5~arcsec$^2$ for the 70~$\mu$m data
and 9~arcsec$^2$ for the 160~$\mu$m data.  The residual backgrounds in
the data were measured in regions outside the optical discs of the
galaxies and subtracted, and then flux calibration factors (given as
$702 \pm 35$ MJy sr$^{-1}$ [MIPS instrumental unit]$^{-1}$ for the
70~$\mu$m data by \citet{getal07} and $41.7\pm5$ MJy sr${^-1}$ [MIPS
  instrumental unit]$^{-1}$ for the 160~$\mu$m data by
\citet{setal07}) were applied to the data.  An additional 70~$\mu$m
nonlinearity correction given as
\begin{equation}
f_{70\mu m}(\mbox{true})=0.581(f_{70\mu m}(\mbox{measured}))^{1.13}
\end{equation}
by \citet{dggetal07} was applied to coarse-scale imaging data where
the surface brightness exceeded 66 MJy sr$^{-1}$. 

The final 70~$\mu$m image is affected by latent images, which are dark
images that appear after the array has observed a bright source.  The
latent image effects result in dark streaks to the northwest and
southeast of the galaxy centre at the edge of the region detected at
70~$\mu$m (at $\alpha=$7:37:20 $\delta=$+65:40:00 and $\alpha=$7:36:35
$\delta=$+65:32:00).  Our algorithms tend to exclude the emission from
these regions, so the latent image effects only have a very minor
impact on the analysis.

We performed several analysis steps in using 70 and 160~$\mu$m flux
densities to calculate dust masses.  First, we convolved the 70~$\mu$m
data with a kernel that matches the PSF of the data to the PSF of the
160~$\mu$m data (see Sec.~\ref{s_data_kernel} below).  Next, we
regridded the 160~$\mu$m data so that its coordinate system matched
the 70~$\mu$m data.  After this step, we only used data for pixels
detected at the $10\sigma$ level in the convolved data for each wave
band so that systematic effects such as offsets in the background
would not strongly affect the results.  We proceeded to fit
blackbodies $B_\nu(T)$ with temperatures $T$ modified by
$\lambda^{-2}$ emissivity functions to the 70 and 160~$\mu$m data for
each pixel meeting this signal-to-noise criterion so as to determine
the approximate dust temperatures.  Following this, we calculated dust
masses within individual pixels using
\begin{equation}
M_{dust} = \frac{D^2 f_{\nu}}{\kappa_{\nu} B_\nu(T)},
\label{e_dustmass}
\end{equation}
which is a variant of an equation derived by \citet{h83}.  In this
equation, $D$ is the distance to the galaxy, $f_{\nu}$ is the flux
density, $\kappa_{\nu}$ represents the absorption opacity of the dust,
and $B_\nu(T)$ is the blackbody function for the best fitting
temperature $T$.  We used the 160~$\mu$m flux densities in
Equation~\ref{e_dustmass} because the data were taken from the
Rayleigh-Jeans side of the thermal dust emission and are therefore
less sensitive to uncertainties in temperature.  The value for
$\kappa_{\nu}$ at 160~$\mu$m is given by \citet{ld01} as
12.0~cm$^2$~g$^{-1}$.  The dust masses were then converted to dust
surface densities by dividing the pixels by their surface area in
pc$^2$.  As a test of the robustness of the results against systematic
offsets in the background, we increased the background in one wave
band by $3\sigma$, decreased it in the other by $3\sigma$, and then
measured the total dust mass and the exponential scale length of the
radial profile for the dust surface density (described in
Section~\ref{s_gasdust}).  We found that the dust mass only varied
$<10$\% and that the scale length varied by only $1\sigma$ when this
was done, even though systematic offsets of this magnitude seem highly
unrealistic.

This is a simplistic approach to calculating dust masses, but it
should provide approximate measurements of the dust masses within this
galaxy.  More sophisticated dust models, such as those presented by
\citet{dh02} and \citet{ddbetal07}, include multiple thermal
components of dust heated by radiation fields with variable strengths
and also account for the different spectral energy distributions of
grains of different sizes.  Unfortunately, those models include many
free parameters, and so when only two wave bands are available for
defining the far-infrared dust SED, as is the case for our analysis of
subregions in NGC~2403, we cannot accurately apply complex dust SED
models.  Fortunately, when the dust masses predicted by the models of
\citet{ddbetal07} are compared with the dust masses calculated using a
single thermal component fit to $\geq 70$~$\mu$m data, the masses agree to
within a factor of 2-4 \citep{rtbetal04, betal06}.  Therefore, the
dust masses estimated from a single thermal component should still be
useful for our analysis.

It is possible that the observations here may have missed the presence
of very cold dust (dust with temperatures $<15$~K) that may not
contribute strongly to the 70 and 160~$\mu$m bands but that may
constitute the bulk of the dust mass within this galaxy.  If present,
this very cold dust emission would be detected at 200-1000~$\mu$m.
Unfortunately, we do not have any useful data with which to search for
the presence of such emission.  Some archival JCMT SCUBA data are
available, but nothing is detected with a signal-to-noise level of
$>3$ in the data.  Moreover, it is unclear whether submillimetre
emission that exceeds thermal dust emission models fit to
$<200$~$\mu$m originates from large masses of cold dust or from dust
with unexpectedly high submillimetre emissivities.  In any case,
most current studies on this subject suggest that the 70 and
160~$\mu$m bands can be used without submillimetre data to produce
reasonably accurate estimates of the dust mass. See
Appendix~\ref{a_dust} for an extended discussion.

\subsection{CO data}

\subsubsection{CO~$J$=(3-2) data}

The CO~$J$=3-2 line observations were obtained at the JCMT as part of
the JCMT NGLS, using the 16 element heterodyne array HARP-B with the
backend spectrometer ACSIS \citep{betal09}.  The FWHM of the beam is
14.5~arcsec.  We give an overview of the observation and data
reduction procedures here.  For a more detailed explanation of the
general JCMT NGLS observing and data reduction process for raster maps
see Warren et al. (2009, submitted).

Observations for NGC~2403 by the JCMT NGLS took place over two runs in
November 2007 and January 2008.  We mapped the galaxy using a
basketweave raster scanning method, with half array steps between scan
rows/columns.  The target area was a rectangular region with a
position angle of $127^\circ$ rotated from north through east and with
a size of 11.7~arcmin~$\times$~6.2~arcmin, which corresponds to half
of the optical diameter along both the major and minor axes.  The
correlator was configured to have a bandwidth of $\sim$1~GHz and a
resolution of 0.488~MHz (0.43~km\,s$^{-1}$ at the frequency of the
$^{12}$CO\,$J$=3-2 transition) centered at 121~km\,s$^{-1}$.  For each
scan, we integrated for 5~s per pointing within the target field, and
scans were repeated until we reached the target RMS for the survey
(19~mK ($\rm T_A^*$) after binning to 20~km~s$^{-1}$ resolution).
After rejecting one half of an individual scan that failed the survey
quality assurance criteria, the total integration time was up to 62.5
seconds per point (depending on the number of contributing receptors
that were not flagged or missing in the array).

Data reduction and most of the analysis were carried out with the
version of the Starlink software package that is maintained by the
Joint Astronomy Centre \citep{cdbjce08}\footnote{Available for
  download from http://www.jach.hawaii.edu.}.  First, data for
individual receptors with bad baselines were flagged and removed from
the data.  Generally, two to four receptors in a scan were flagged.
After this, the raw scans were combined into a data cube using a ${\rm
  sinc}(\pi x){\rm sinc}(k\pi x)$ kernel as the weighting function to
determine the contribution of individual receptors to each pixel.  The
${\rm sinc}(\pi x){\rm sinc}(k\pi x)$ function was used instead of
other weighting functions because it significantly reduces the noise
in the maps while having a very minor effect on the resolution. The
resulting cube was then trimmed to remove the leading and trailing
scan ends outside the target field, where the scan coverage is
incomplete.  A mask was created to identify line-free regions of the
data cube, and a third-order baseline was fit to those line-free
regions.  The clumpfind algorithm \citep{wdb94}, implemented as part
of the Starlink/CUPID task findclumps \citep{brje07}, was used to
identify regions with emission above two times the rms noise in a data
cube that had been boxcar smoothed by 3 pixels and 25 velocity
channels.  Moment maps were created from the original data cube using
a mask created from findclumps.  We converted the CO~$J$=(3-2) data
from corrected antenna temperature ($\rm T_A^*$) to the main beam
temperature scale by dividing by $\eta_{MB}=0.60$.  The CO~$J$=(3-2)
emission is not detected beyond radii of $\sim3.75$~kpc (which
corresponds to an angular scale of $\sim250$~arcsec along the major
axis).

For some of the analysis in this paper, the CO~$J$=(3-2) emission is
converted into molecular gas surface density.  The conversion factor
is dependent on a comparison between the CO~$J$=(3-2) and $J$=(1-0)
line emission.  We discuss both of these topics in
Section~\ref{s_cocompare}.

\subsubsection{CO~$J$=(1-0) data}
\label{s_data_co_co10}

We compared the CO~$J$=(3-2) data with CO~$J$=(1-0) data from two
sources.  We primarily used CO~$J$=(1-0) data taken by \citet{tw95}
using the National Radio Astronomy Observatory (NRAO) 12 meter
telescope.  The observations were performed as a series of pointings
separated by 30~arcsec that approximately cover the central
5.5~arcmin~$\times 4$~arcmin of the galaxy.  These data have a FWHM of
54~arcsec and a spectral resolution of 2.6~km~s$^{-1}$.  Since the
spacing between the pointings is smaller than the beam, we can treat
these data as a fully sampled map.  Table~1 by \citet{tw95} gives the
integrated line intensities (based on main beam temperatures) and
uncertainties that they measured in individual pointings across the
optical disc of the galaxy; we used these data to construct a
CO~$J$=(1-0) image.  Upper limits were treated as 0 for our analysis.
See \citet{tw95} for additional details on the observations and data
reduction.

We also had access to CO~$J$=(1-0) data taken in the BIMA Survey of
Nearby Galaxies \citep[SONG; ][]{htrwsvbb03}.  While the $6.3 \times
5.8$~arcsec FWHM of the beam for the data is superior to that for both
HARP-B and the NRAO 12~m data, the BIMA SONG data have problems that
make it less suitable for our analysis.  A comparison of the BIMA SONG
data to the NRAO 12~m data revealed that only $\sim60$\% of the flux
was recovered \citep{htrwsvbb03}, and the area in which sources are
detected in the BIMA SONG data is significantly smaller than that for
either the HARP-B or NRAO 12~m data.  While we still use the BIMA SONG
data for a qualitative comparison to other data, we will not use it
for quantitative comparisons.

\subsection{H{\small I} data}

The H{\small I} data used for this analysis were taken by THINGS
\citep{wbbbktl08}.  The observations were performed with the Very
Large Array in the B, C, and D configurations, which ensured fairly
uniform sensitivity from the largest scales all the way to the
resolution limit set by the long baselines of B-array.  Calibrated,
continuum-subtracted data cubes were created using the Astronomical
Image Processing System.  The uncertainties are dominated by the
calibration uncertainty, which is 5\%.

THINGS released two data products that differ based on the weighting
of the data.  We used the data with the natural weighting.  In these
data, all visibility points in the $uv$ plane are weighted equally.
Large-scale structures, including the diffuse, extended H{\small I}
emission that may be associated with diffuse interstellar dust, should
be more easily detected in the data with the natural weighting.  The
FWHM of the PSF for the naturally weighted data is $8.75 \times
6.75$~arcsec.  According to \citet{wbbbktl08}, a comparison between
the VLA data and single-dish data from the literature compiled by
\citet{ptbgchp03} showed a strong agreement for most THINGS galaxies
including NGC~2403, indicating that significant emission is not
missing from the VLA data because of gaps in the coverage of the $uv$
plane.  See \citet{wbbbktl08} for additional details on the
observations and data reduction as well as the reliability of the
data.  To convert the surface brightnesses $S_\nu$ to atomic gas
surface densities $\sigma_{dust}$ on a pixel-by-pixel basis, we used
the equation
\begin{equation}
\begin{array}{l}
\frac{\sigma_{dust}}{\mbox{M}_\odot \mbox{pc}^{-2}} = \\
   8870
   \left( \frac{S_\nu}{\mbox{Jy beam}^{-1} \mbox{km s}^{-1}} \right)
   \left( \frac{\mbox{arcsec}^2}{\mbox{FWHM}_{maj}\mbox{FWHM}_{min}} \right)
\end{array}
\end{equation}
based on equations given by \citet{wbbbktl08}.

\subsection{Convolution kernels}
\label{s_data_kernel}

The various images described above have PSFs with varying widths, and
the shapes of the PSFs vary as well.  In particular, the {\it
Spitzer} PSFs have very pronounced Airy rings whereas most other
PSFs can be approximated as Gaussian functions.  To properly compare
the data to each other, we need convolution kernels that not only
match the FWHMs of the PSFs to each other but also match the shapes to
each other.

We created kernels for this purpose following the instructions given
by \citet{getal08}. We first generated images that represented the
PSFs in the individual wave bands.  For the CO~$J$=(3-2),
CO~$J$=(1-0), and H{\small I} data, we used Gaussian functions.  For
the 3.6 and 8~$\mu$m data, we used PSFs produced with
STinyTim\footnote{\raggedright Available from
  http://ssc.spitzer.caltech.edu/archanaly/contributed/browse.html.},
a PSF simulator designed for {\it Spitzer} \citep{k02}.  For the 70
and 160~$\mu$m data, we used empirically-determined PSFs created from
extragalactic observations of point-like extragalactic sources
\citep[see][for additional information]{ybl09}.  Table~\ref{t_psf}
gives a summary of these PSFs.

\begin{table}
\begin{center}
\caption{PSFs Used for Creating Convolution Kernels\label{t_psf}}
\begin{tabular}{@{}cccl@{}}
\hline
Telescope &             Wavelength &      FWHM &       Source\\
/ Instrument &          &                 (arcsec) &   \\
\hline
{\it Spitzer} / IRAC &  3.6~$\mu$m &      1.7 &        Model (STinyTim)\\
&                       8.0~$\mu$m &      2.0 &        Model (STinyTim)\\
{\it Spitzer} / MIPS &  70~$\mu$m &       18 &         Empirical\\
&                       160~$\mu$m &      38 &         Empirical\\
JCMT / HARP-B &         CO~$J$=(3-2) &    14.5 &       Gaussian function\\
NRAO 12~m &             CO~$J$=(1-0) &    54 &         Gaussian function\\
VLA &                   H{\small I} &     $8.75 \times 6.75$ &
          Gaussian function$^a$\\
\hline
\end{tabular}
\end{center}
$^a$ The position angle of the major axis is $25.2^\circ$ from north to east.
\end{table}

We then used the equation
\begin{equation}
K(x,y)=F^{-1} \left[ \frac{W(\omega)F[PSF_2(x,y)]}{F[PSF_1(x,y)]} \right]
\end{equation} 
to create convolution kernels $K(x,y)$ that match PSF$_1$ to PSF$_2$.
F is a Fourier transform, $W(\omega)$ is a radial Hanning
truncation function given by
\begin{equation}
W(\omega) = \left\{ \begin{array}{cr} 
\frac{1}{2} \left[ 1+\cos \left( \frac{2\pi\omega}{\omega_0} \right) \right]
    & \omega \leq \omega_0 \\
0 & \omega > \omega_0 \end{array} \right. 
\end{equation} 
that is used to suppress high-frequency spatial noise in the resulting
kernels.  For each kernel, we adjusted $\omega_0$ to produce kernels
which, when visually inspected, contained as much structure related to
the PSF transformation as possible while excluding high frequency
noise.  However, for converting from one idealized Gaussian profile to another,
we created convolution kernels using the equation
\begin{equation}
K(r)=e^{-r^2/2(\sigma_2^2-\sigma_1^2)}
\end{equation} 
where $\sigma_1$ is the width of the narrower Gaussian function that
is being blurred to match the width $\sigma_2$ of the broader Gaussian
function.

The final kernels that we created convert data to match two different
PSFs.  For comparing PAH emission to atomic or molecular gas, the
kernels match the data to the PSFs of the CO~$J$=(3-2) data.  For
comparing gas and dust emission, the kernels match the data to the
PSFs of the 160~$\mu$m data.  For the comparison between the
CO~$J$=(3-2) and $J$=(1-0) data, the PSFs of the data were matched to
the PSF of the NRAO data.  Note that we used a pre-made kernel created
by K. D. Gordon for matching the PSF of the 70~$\mu$m data to that of
the 160~$\mu$m data\footnote{\raggedright The kernel is
  available at
  http://dirty.as.arizona.edu/$\sim$kgordon/mips/conv\_psfs/conv\_psfs.html.}.
The convolution kernels were applied to the CO~$J$=(3-2) data before
spectral line extraction.

\subsection{Abundance information}
\label{s_data_abundance}

Since metallicity has been shown to have a possible effect on
the CO to H$_2$ conversion factor (discussed further in
Section~\ref{s_cocompare}) and since metallicity variations could explain
variations in gas-to-dust ratios (discussed further in
Section~\ref{s_gasdust}), we need abundance gradient information for our
analyses.  Many authors have measured and published gradients in
12+log(O/H) within this galaxy, including \citet{ftp86}, \citet{ve92},
\citet{zkh94}, \citet{gsssd97} and \citet{vetal98}.  After scaling the
various gradients so that they correspond to a galaxy distance of
3.13~Mpc, we find that the 12+log(O/H) gradients vary from
$-0.0774\pm0.0014$~dex~kpc$^{-1}$ \citep{vetal98} to
$-0.098\pm0.009$~dex~kpc$^{-1}$ \citep{gsssd97} with a mean of
-0.084~dex~kpc$^{-1}$ and a standard deviation of
0.009~dex~kpc$^{-1}$.  The value of 12+log(O/H) at the centre of the
galaxy is also given by \citet{ve92}, \citet{zkh94},
and \citet{vetal98}; the mean of their measurements is 8.4 with a
standard deviation of 0.13.  We use these values for the abundance
gradients in NGC~2403.  Compared to the abundance gradients for other
galaxies reported by \citet{ve92} and \citet{zkh94}, both of whom
worked with relatively large samples, the abundance gradients measured
in NGC~2403 are rather typical.

Although the studies cited above report consistent 12+log(O/H)
gradients for NGC~2403, more recent observations of abundance
gradients in other galaxies have shown that the methods used in the
above references may contain systematic errors \citep[e.g.]{p01,
  bgk04, rs08}.  Moreover, newer methods for calculating oxygen
abundances show that the gradients may be significantly shallower.
However, because no abundance gradient data for NGC~2403 has been
published using these newer techniques, we will use these older
abundance gradients for now but discuss how shallower abundance
gradients could affect the results where appropriate.

\section{Images}
\label{s_img}

\begin{figure*}
\begin{center}
\epsfig{file=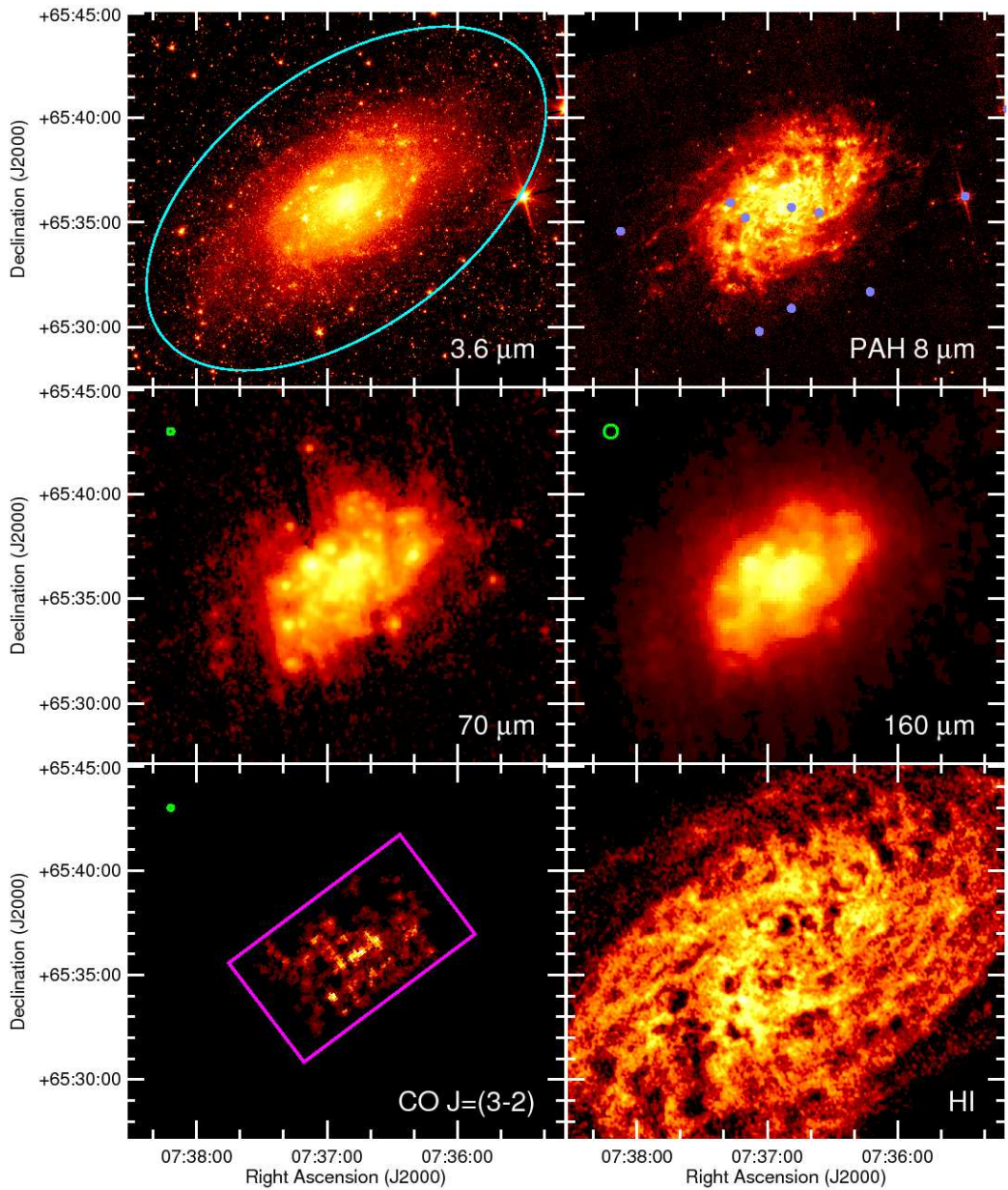}
\end{center}
\caption{Images of the entire optical disc of NGC~2403 in multiple
  wave bands.  The 3.6~$\mu$m image traces starlight.  The PAH
  8~$\mu$m image traces primarily PAH emission.  The 70 and 160~$\mu$m
  images trace cold dust emission.  The CO~$J$=(3-2) image shows CO
  emission associated with molecular gas.  The H{\small I} image shows
  atomic gas emission.  All images are 21~arcmin~$\times 18$~arcmin
  with north up and east to the left.  Logarithmic colour scales are
  used for all images except the 160~$\mu$m image, where the
  logarithmic scale to the second power was used to enhance the
  structure in the image, and the CO~$J$=(3-2) image, where a linear
  colour scale was used.  The colour scales were selected so as to
  best enhance the structures visible in the optical disc of the
  galaxy.  Green circles in the upper left corners of the images show
  the beam size when the FWHM is greater than 10~arcsec; smaller beam
  sizes are difficult to illustrate in this image.  The optical disc
  of the galaxy as defined by \citet{ddcbpf91}, is shown as a cyan
  ellipse in the 3.6~$\mu$m image.  The blue circles in the PAH
  8~$\mu$m image show the positions of foreground stars that were
  brighter than or as bright as the continuum emission from the galaxy
  at 8~$\mu$m.  These stars were masked out in the analysis.  The
  magenta box in the CO~$J$=(3-2) image shows the
  region observed with HARP-B.  Pixels with non-detections in the
  CO data are set to black.}
\label{f_img}
\end{figure*}

\begin{figure*}
\begin{center}
\epsfig{file=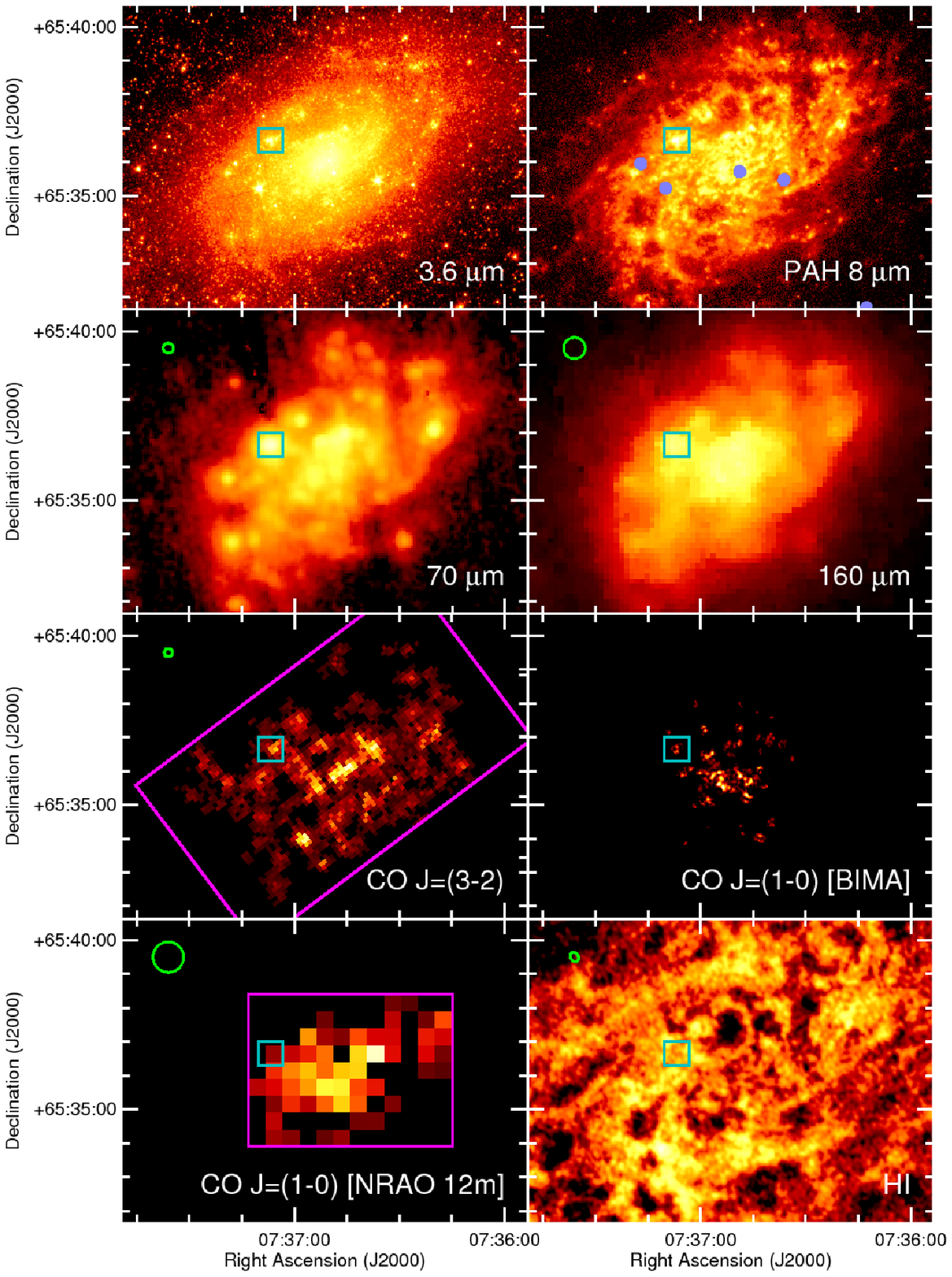}
\end{center}
\caption{Images of the inner $12\times9$~arcmin of NGC~2403 in
  multiple wave bands.  The 3.6~$\mu$m, 8.0~$\mu$m, 70~$\mu$m,
  160~$\mu$m, CO~$J$=(3-2), and H{\small I} images are the same as the
  images in Figure~\ref{f_img} except that the images are cropped and
  that the colour scales have been adjusted to show greater contrast
  in the centre.  The colours in the two additional CO~$J$=(1-0)
  images (from the BIMA SONG survey and from the NRAO 12~m observations
  by \citet{tw95} are scaled linearly.  The magenta box in the
  CO~$J$=(1-0) image from the NRAO 12~m shows the region that was
  covered in the observations.  Pixels with non-detections in the CO
  data are set to black.  The cyan square indicates the location of VS
  44, which is region 44 in the catalogue by \citep{vs65} and region
  128 in the catalogue by \citet{hk83}.  See the caption for
  Figure~\ref{f_img} for additional information on the other symbols
  and lines in the figure.}
\label{f_img_small}
\end{figure*}

Figure~\ref{f_img} shows in their native resolution the 3.6~$\mu$m,
8.0~$\mu$m, 70~$\mu$m, 160~$\mu$m, CO~$J$=(3-2), and H{\small I}
images for the entire optical disc of the galaxy.  In addition,
Figure~\ref{f_img_small} shows the inner $12\times9$~arcmin for all of
these images as well as the CO~$J$=(1-0) images from the NRAO~12~m
telescope and from the BIMA SONG survey.  In interpreting these
images, we will refer to the H{\small II} region marked with the cyan
square in Figure~\ref{f_img_small} as VS 44.  This is region 44 in the
catalogue produced by \citep{vs65}, and it corresponds to region 128 in
the atlas of H{\small II} regions mapped by \citet{hk83}.  VS 44 is
the brightest source of both H$\alpha$ emission \citep{drms99} and
24~$\mu$m emission \citep{betal08} and therefore is the site with the
strongest star formation activity in the galaxy.  Since the region is
located well outside the nucleus, it can be used to distinguish
between effects related to radius and effects related to star
formation activity or infrared surface brightness.

The stellar emission, which is shown by the 3.6~$\mu$m image, looks
similar to what is expected for a late-type spiral galaxy; the bulge
is close to non-existent, and the disc, while well-defined, has a
clumpy structure.  The PAH 8~$\mu$m image shows the clumpy dust
structures in the ISM.  Some of this structure is also visible in the
70 and 160~$\mu$m images despite the lower resolution of the data.
The dust structure is fairly typical compared to most late-type spiral
galaxies \citep{betal07}.  The CO~$J$=(3-2) and $J$=(1-0) images all
appear similar to each other.  Some of the structures seen in the CO
emission, such as the spiral arm structure to the west of the centre
and a few CO-bright regions like VS 44, are similar to those traced by
PAH and longer-wavelength dust emission.  However, the CO maps contain
a notable hole close to the centre of the optical disc, whereas PAH
8~$\mu$m and longer-wavelength dust emission is still present in this
region.  Interestingly, this hole corresponds to the peak in diffuse
X-ray emission found by \citet{fcso02}.  We discuss this further in
Sections~\ref{s_gasdust} and \ref{s_pahco}.  In contrast to all of
these images, the H{\small I} emission is much more extended; the
emission is detectable well outside the optical disc of the galaxy.  A
number of shell-like structures are visible within the H{\small I}
data.  These shells were also identified by \citet{tbw98}, who
associated them with star-forming regions.  While bright objects in
many of the other wave bands, such as VS 44, do not have H{\small I}
counterparts, some of the shells in the H{\small I} image, such as the
ones at $\alpha=$7:37:15 $\delta=$+65:34:30 and $\alpha=$7:36:50
$\delta=$+65:37:30, do have counterparts in other wave bands.

\begin{figure}
\begin{center}
\epsfig{file=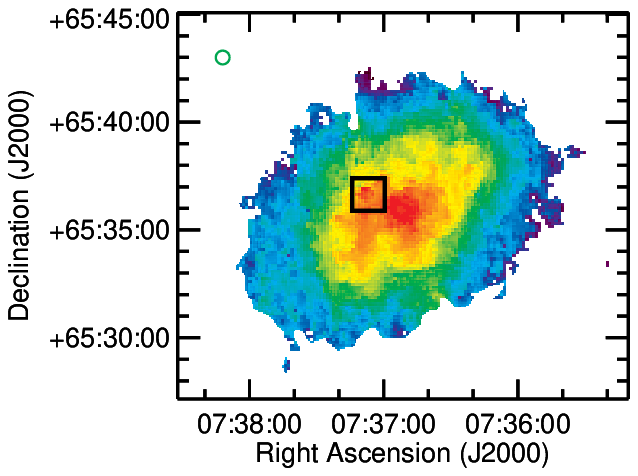}
\epsfig{file=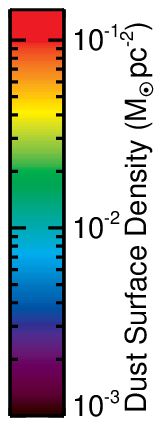}
\end{center}
\caption{Image of the dust surface density in NGC~2403 produced after
  the analysis in Section~\ref{s_data_dust}.  The surface densities
  were corrected for the inclination of this galaxy (given as
  $62.9^\circ$ by \citet{dwbtok08}).  The image size and orientation is the
  same as the size and orientation of the images in
  Figure~\ref{f_img}.  The green circle in the upper left corner shows
  the 38~arcsec FWHM of the PSF, which is equivalent to the PSF of the
  160~$\mu$m data.  The uncertainties in the data are $\sim20$\%.  The
  black square indicates the location of VS 44.}
\label{f_img_dustmass}
\end{figure}

\begin{figure}
\begin{center}
\epsfig{file=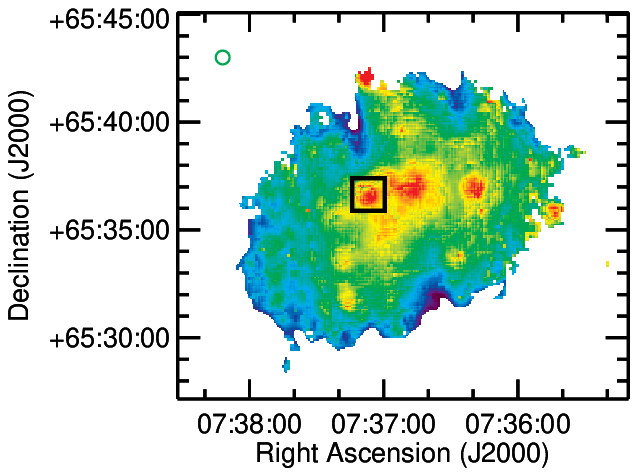}
\epsfig{file=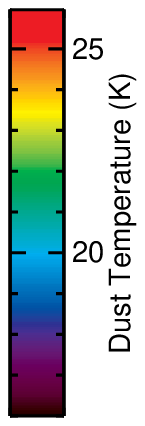}
\end{center}
\caption{Image of the dust temperatures in NGC~2403 produced after the
  analysis in Section~\ref{s_data_dust}.  The image size and
  orientation is the same as the size and orientation of the images in
  Figure~\ref{f_img}.  The green circle in the upper left corner shows
  the 38~arcsec FWHM of the PSF, which is equivalent to the PSF of the
  160~$\mu$m data.  The uncertainties are $\ltsim1$~K.  The black
  square indicates the location of VS 44.  Regions with
  temperatures of $<20$~K at $\alpha=$7:37:20 $\delta=$+65:40:00 and
  $\alpha=$7:36:35 $\delta=$+65:32:00 may have been strongly affected
  by latent image effects.}
\label{f_img_dusttemp}
\end{figure}

Figure~\ref{f_img_dustmass} and \ref{f_img_dusttemp} show the dust
surface density and temperature maps derived as part of the analysis
in Section~\ref{s_data_dust}.  The dust mass image exhibits structures
that appear similar to those seen at 70 and 160~$\mu$m.  However,
star-forming regions appear notably enhanced in the temperature map.
The dust temperatures range from 26~K in infrared-bright regions
(including VS 44) to 17~K in the fainter diffuse regions.
Note that the $<20 K$ regions located at approximately
$\alpha=$7:37:20 $\delta=$+65:40:00 and $\alpha=$7:36:35
$\delta=$+65:32:00 may have been affected by latent image effects in
the 70~$\mu$m band.  The median temperature is 22~K.  These
temperatures are similar to those measured in other nearby galaxies
when single thermal components have been fit to far-infrared ISO or
{\it Spitzer} data that sample the peak of the thermal dust emission
from 60 to 200~$\mu$m \citep[e.g.][]{ptvpm02, betal03, eetal04,
  rtbetal04, betal06, petal06}.

\section{The CO~$J$=(3-2)/$J$=(1-0) ratio and the conversion
from line intensity to molecular gas surface density}
\label{s_cocompare}

Given the superior resolution of the HARP-B CO~$J$=(3-2) data compared
to the older NRAO 12~m CO~$J$=(1-0) data and given the superior
spatial coverage of the HARP-B data to both the NRAO 12~m and BIMA
SONG data, it would be preferable to use the HARP-B data to calculate
molecular gas surface densities for comparisons to dust emission,
particularly PAH 8~$\mu$m emission.  However, the conversion from CO
line emission to molecular gas mass ($X_{CO}$) is generally based on
the CO $J$=(1-0) line.  Therefore, the CO $J$=(3-2)/$J$=(1-0) ratio
needs to be determined to convert the CO~$J$=(3-2) line to a molecular
gas surface density.

\begin{figure}
\begin{center}
\epsfig{file=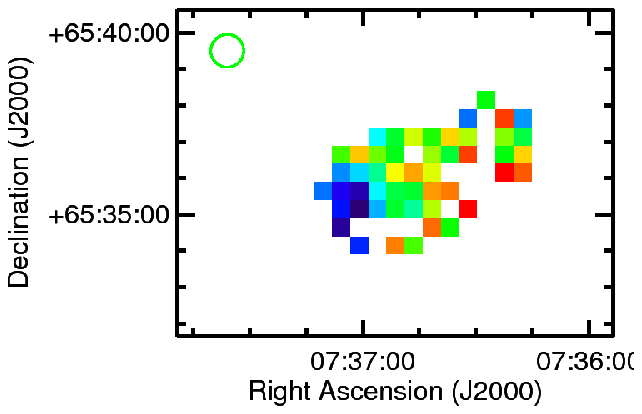}
\epsfig{file=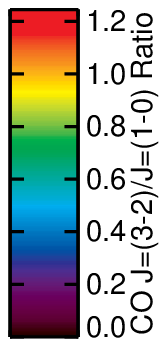}
\end{center}
\caption{Image of the ratio of the CO~$J$=(3-2)/$J$=(1-0) line ratio
  for the central $12\times9$~arcmin of NGC~2403.  The orientation,
  size, and central coordinates of the image have been selected to
  match those of Figure~\ref{f_img_small}.  The 30~arcsec pixels
  correspond to the positions of the pointed observations performed by
  \citet{tw95}.  To make this plot, the PSF of the data in the
  CO~$J$=(3-2) data cube were matched to the PSF of the CO~$J$=(1-0)
  data, the CO~$J$=(3-2) line intensities were measured, and then the
  CO~$J$=(3-2) data were rebinned to match the pixel scale of the
  CO~$J$=(1-0) data.  The green circle in the upper left corner shows
  the 54~arcsec PSF of the CO~$J$=(1-0) data.  The uncertainties in
  this ratio are mainly dependent on the uncertainties in the
  CO~$J$=(1-0) data; the typical signal-to-noise values of the ratios
  range from $3\sigma$ to $8\sigma$.  Locations where line emission
  was not measured above the $3\sigma$ level and locations affected by
  artifacts along the edge of the region observed in the CO~$J$=(3-2)
  line were left blank.}
\label{f_img_co32vsco10}
\end{figure}

For this analysis, we compared the HARP-B CO~$J$=(3-2) data to the
NRAO 12~m CO~$J$=(1-0) data from \citet{tw95} for the reasons
discussed in Section~\ref{s_data_co_co10}.  We examined both median
line intensities and intensities measured in the pointing positions
used by \citet{tw95}.  For these comparisons, we use intensities based
on the main beam temperatures.  For comparing median line intensities,
we used CO~$J$=(3-2) data where the resolution and pixels are matched
to those of the CO~$J$=(1-0) data, and we only selected pixels that
are $3\sigma$ detections in both images.  The median CO~$J$=(1-0) line
intensity in these pixels is 1.56~K~km~s$^{-1}$, while the median
CO~$J$=(3-2) intensity is 1.08~K km s$^{-1}$.  This gives a median
CO~$J$=(3-2)/$J$=(1-0) line ratio of 0.69.  The standard deviation of
the ratio across the disc is 0.29.  Figure~\ref{f_img_co32vsco10}
shows how the line ratio varies across the centre of the galaxy.  The
ratio does not appear to exhibit any clear structure, although the
ratio is low in the southeast side of the region, which corresponds to
a location where the CO~$J$=(3-2) ratio is also low.  The ratios range
from $0.11 \pm 0.3$ to $1.2 \pm 0.3$.

For comparison, \citet{tw94} and \citet{wwt97} measured ratios for
individual giant molecular clouds in M33 ranging from 0.4 to 0.8, and
\citet{mhws99} measured CO~$J$=(3-2)/$J$=(1-0) ratios in 28 nearby
spiral galaxies ranging from 0.2 to 0.7.  The median
CO~$J$=(3-2)/$J$=(1-0) ratio for NGC~2403 appears high but typical
compared to these other measurements.  However, the factor of 10
variation in the CO~$J$=(3-2)/$J$=(1-0) ratio across the disc of
NGC~2403 is high compared to what was measured by these other surveys,
and the minimum and maximum fall outside the typical ranges of the
objects in the other surveys.  It is quite possible that the
variations in the ratio that we observed in NGC~2403 are real and that
such variations have not been observed in other surveys either because
the the regions were unresolved or undetected.  \citet{wetal09} using
CO~$J$=(3-2) data from the JCMT NGLS found evidence for
CO~$J$=(3-2)/$J$=(1-0) ratios that vary by a factor of 10 in NGC~4569
as well.  It could be possible that further analysis of the JCMT NGLS
data will reveal that such variations are actually typical for nearby
galaxies.

\begin{figure*}
\begin{center}
\epsfig{file=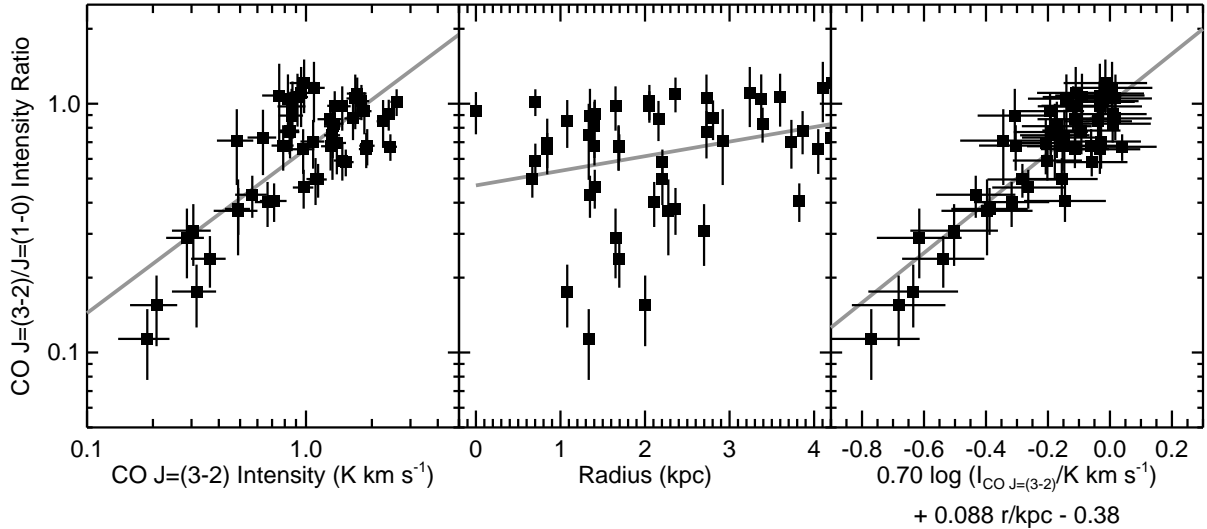}
\end{center}
\caption{The CO~$J$=(3-2)/$J$=(1-0) intensity ratio plotted as a
  function of CO~$J$=(3-2) intensity (left panel), radius (center
  panel), and the right hand side of Equation~\ref{e_corelation}
  (right panel).  The data points are for the 30~arcsec square regions
  shown in Figure~\ref{f_img_co32vsco10}.  The grey lines are the best
  fit lines to the data.}
\label{f_cocompare}
\end{figure*}

Given the strong variations in the ratio, we did not feel confident
about applying a simple scaling term to the CO~$J$=(3-2) data to scale
it to match the CO~$J$=(1-0) data.  We therefore examined how the
CO~$J$=(3-2)/$J$=(1-0) ratio could be derived from either the radius
or the CO~$J$=(3-2) surface brightness.  For this analysis, we used
the CO~$J$=(3-2) and $J$=(1-0) data that were prepared for
Figure~\ref{f_img_co32vsco10}.  A fit between
log(CO~$J$=(3-2)/$J$=(1-0)) and a linear function of both
log(CO~$J$=(3-2)/(K km s$^{-1}$)) and radius (calculated using the
inclination of $62.9^|circ$ given by \citet{dwbtok08}) resulted in the lowest
reduced $\chi^2$ value.  The relation is
\begin{equation}
\begin{array}{l}
\log\left(\frac{I_{CO ~ J=(3-2)}}{I_{CO ~ J=(1-0)}}\right) 
= (0.70 \pm 0.07)\log\left(\frac{I_{CO ~ J=(3-2)}}
{\mbox{K km s}^{-1}}\right) \\
+ (0.088 \pm 0.014)\left(\frac{r}{\mbox{kpc}}\right) - (0.38 \pm 0.03).
\end{array}
\label{e_corelation}
\end{equation}
Figure~\ref{f_cocompare} shows the relation between the
CO~$J$=(3-2)/$J$=(1-0) intensity ratio and CO~$J$=(3-2) intensity, the
relation between the intensity ratio and radius, and the relation
given by Equation~\ref{e_corelation}.  The dependence of the ratio on
CO~$J$=(3-2) intensity is very apparent, but the dependence of the
ratio on radius is fairly weak.  Nonetheless, including a term for
radius in the conversion between the CO~$J$=(3-2) intensity and the
CO~$J$=(3-2)/$J$=(1-0) ratio does provide a better fit (in terms of
the reduced $\chi^2$) than a function that does not include such a
term.  We also found that the Pearson's correlation coefficient
increased from 0.54 for the data in the left panel of
Figure~\ref{f_cocompare} to 0.81 for the data in the right panel,
confirming that the extra radial distance term reduced the scatter.
Based on the scatter around the relation given in
Equation~\ref{e_corelation}, the conversion from the CO~$J$=(3-2)
intensity to the CO~$J$=(1-0) intensity should be accurate to within
20\%.

Given these results, we can now scale the CO~$J$=(1-0) to H$_2$
conversion factor for use with the CO~$J$=(3-2) line.  $X_{CO}$ for
the $J$=(1-0) transition may depend on several environmental factors.
In particular, it has been shown that it may be dependent on
metallicity \citep[e.g.][]{w95, i97b, i00, bhhfk00, ibrsw03, smrdd04,
  i05}.  We have decided to use the conversion
\begin{equation}
\log(X_{CO ~ J=(1-0)})=12.2 - 2.5 \log[\mbox{O}/\mbox{H}]
\label{e_xco_metal}
\end{equation}
proposed by \citet{i00}, which is based on determining the $X_{CO ~
  J=(1-0)}$ needed to keep the ratio of total gas surface density to
far-infrared surface brightness constant in CO clouds in the Large and
Small Magellanic Clouds.  This equation is also found to be in
agreement with measurements based on gamma-ray emission from the Milky
Way \citep{smrdd04}.  We use this equation along with the abundance
gradients given in Section~\ref{s_data_abundance} in one set of
calculations of the molecular gas surface densities.  

However, recent observational results in which the CO velocity
dispersions, radii, and CO luminosities of giant molecular clouds are
used to infer their mass and hence $X_{CO ~ J=(1-0)}$ suggest that the
conversion factor does not vary with metallicity \citep{bfklmr07, blrwb08},
contradicting the results derived using other methods.  Therefore, we
also use a second set of molecular gas surface density data calculated
using $X_{CO ~ J=(1-0)} = 1.9 \times 10^{20}$ cm$^{-2}$ (K km
s$^{-1}$)$^{-1}$, which was derived by \citet{sm96} using models of
gamma ray scattering that did not include radial variations.

\citet{i97b} indicated that radiation field strength could be a second
factor that affects $X_{co}$, although some authors have not found
evidence for such a dependency \citep[e.g.][]{blrwb08}.  The
application of the relation found by \citet{i97b} to our data does not
appear straightforward.  The metric for radiation field intensity used
by \citet{i97b} uses the ratio of the far-infrared surface
brightnesses measured using IRAS data to the neutral hydrogen density.
Because IRAS data does not sample the Rayleigh-Jeans side of the dust
SED in the same way that {\it Spitzer} data does, we would need to
convert the far-infrared fluxes derived from {\it Spitzer} data to
match that calculated from IRAS data, which may not be
straightforward.  Moreover, \citet{i97b} derived his relation using
subregions within dwarf galaxies where radial variations in abundances
were not present and therefore where variations in the gas-to-dust
ratio should not be present, whereas NGC~2403 clearly exhibits radial
variations in abundances and therefore may exhibit radial variations
in the gas-to-dust ratio that would affect the radiation field
intensity metric.  Fortunately, \citet{i97b} found that it should
still be possible to derive a reasonable approximation for $X_{CO}$
based on its relation to oxygen abundances alone.  Therefore, for
simplicity, we will not attempt to include an additional correction to
$X_{CO}$ for radiation field intensity, although we still discuss the
implications of this dependency when appropriate.

\section{Comparison of atomic gas, molecular gas, and dust surface densities}
\label{s_gasdust}

To compare the molecular, atomic, and dust surface densities, we first
measured the radial profiles of the surface densities as well as the
gas-to-dust ratios.  First, all data were convolved with kernels that
match the PSFs to that of the 160~$\mu$m data before radial profiles
were extracted.  The gas-to-dust ratio was calculated on a
pixel-by-pixel basis from images regridded to match the astrometry of
the dust surface density images.  We then measured means, standard
deviations, and uncertainties in mean values within elliptical annuli
with widths along the major axes that were equal to 2 pixels (or
9~arcsec in the dust and gas-to-dust ratio data, 14.6~arcsec in the
molecular gas data, and 2~arcsec in the atomic gas data).  This method
for setting the annuli widths was used to avoid problems with noise
caused by having too few pixels fall within given annuli, but the
resulting annuli are small enough that the data, which have a PSF with
a FWHM of 38~arcsec, will be Nyquist-sampled. Pixels where CO or
H{\small I} line emission was not detected were set to 0 in those
data.  The ratio of the axes in the elliptical annuli were set to
account for the inclination of the galaxy \citep[$62.9^\circ$;
][]{dwbtok08}.

\begin{figure}
\begin{center}
\epsfig{file=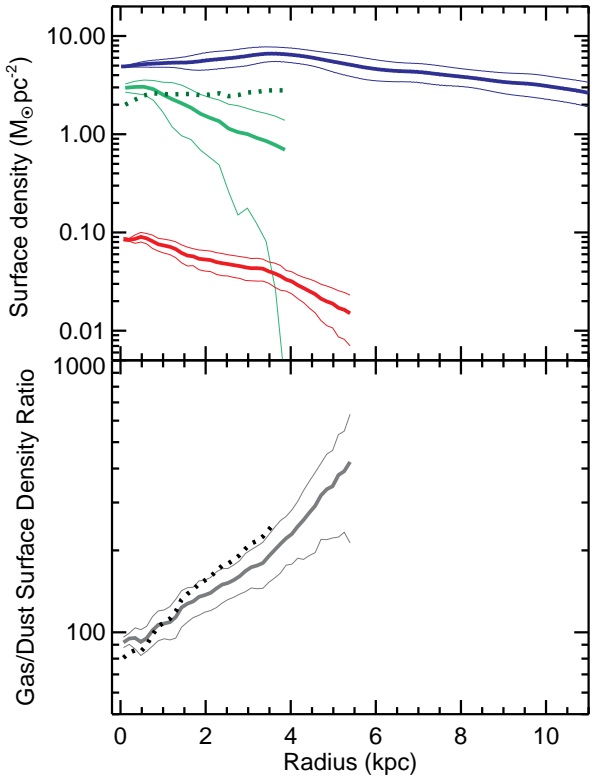}
\caption{Radial profiles of the H{\small I} surface density (solid
  blue line, top panel), H$_2$ surface density (green lines, top
  panel), and dust surface density (solid red line, top panel) as well
  as the total (molecular and atomic) gas-to-dust surface density
  ratio (bottom panel) plotted up to 10~kpc, which is equivalent to
  the edge of the optical disc.  For the H$_2$ and gas/dust ratios,
  the solid lines are for quantities calculated using a constant
  $X_{CO}$ and the dotted lines are for quantities calculated using a
  value of $X_{CO}$ that depends on 12+log(O/H).  The standard
  deviations of the data are plotted as thinner lines around all
  radial profiles except for the H$_2$ surface density and gas-to-dust
  ratios calculated with a variable $X_{CO}$, which have standard
  deviations that are similar to the equivalent quantities calculated
  with the constant $X_{CO}$.  The uncertainties in the mean values
  are generally smaller than the thicknesses of these lines. The H$_2$
  and dust radial profiles are truncated at the maximum radii where
  the quantities could be reliably measured along the major axis (4
  and 5.5~kpc, respectively).  The gas-to-dust ratio calculated using
  the constant $X_{CO}$ is also truncated at the maximum radius at
  which the dust surface densities could be measured.  The gas-to-dust
  ratio calculated using the variable $X_{CO}$ is truncated at 4~kpc
  to show how the change in $X_{CO}$ affects this ratio; beyond this
  radius, the radial profile merges back into the other gas-to-dust
  radial profile.}
\label{f_radpro_gasdust}
\end{center}
\end{figure}

Figure~\ref{f_radpro_gasdust} shows the radial profiles of the
H{\small I}, H$_2$, and dust surface densities.  Two versions of the
H$_2$ surface densities are shown.  The values shown with a solid line
were calculated using a constant $X_{CO}$, and the values shown with
the dotted line were calculated using the formula for $X_{CO}$ given
in Equation~\ref{e_xco_metal}.  For simplicity, we will refer to the
former as the constant $X_{CO}$ molecular gas surface density and the
latter as the variable $X_{CO}$ molecular gas surface density.
Although the standard deviation in the CO~$J$=(3-2) radial profiles
approaches the same level as the mean values at large radii, the
uncertainties in the mean values are no wider than the thicknesses
of these lines.  Hence, the mean values are very well defined even
though significant variance may be seen in CO intensities at specific
radii.

The figure shows that most of the mass in the interstellar medium of
this galaxy is traced by the H{\small I} emission, a result also
obtained by \citet{tw95}.  The H{\small I} surface density is
relatively flat; it increases slightly towards 4~kpc and then declines
gradually beyond that radius.  The constant $X_{CO}$ molecular gas
surface density decreases exponentially with radius.  The scale length
of the constant $X_{CO}$ molecular gas radial profile is $2.32 \pm
0.07$~kpc.  The variable $X_{CO}$ molecular gas surface density,
however, is so close to flat that it is not possible to accurately fit
an exponential function to the profile.  The dust surface density,
like the constant $X_{CO}$ molecular gas surface density, decreases
with radius.  Its radial profile has an exponential scale length of
$3.16 \pm 0.10$~kpc.  If we applied an additional correction for
radiation field strength in $X_{CO}$, then this would increase the
molecular gas surface density in regions with high radiation field
intensities, which would correspond to regions in the centre of
NGC~2403 with higher dust temperatures shown in
Figure~\ref{f_img_dusttemp}.  Therefore, the molecular gas surface
density radial profile would fall in between those of the constant
$X_{CO}$ and variable $X_{CO}$ molecular gas surface densities plotted
in Figure~\ref{f_radpro_gasdust}.

Of particular interest is how the radial profiles of the molecular gas
changes when the abundance-dependent correction is introduced.  If the
variable $X_{CO}$ molecular gas surface density is an accurate
depiction of the true surface density, then it implies that $\sim25$\%
of the ISM at large radii is comprised of molecular gas.  If true,
this would contradict the typical depiction of how the ratio of
molecular to atomic gas varies within spiral galaxies, with the
centres of galaxies being dominated by molecular gas and the outer
discs being dominated by atomic gas \citep[e.g.]{wb02, blwbdmt08}.
However, these prior results rely upon the assumption that $X_{CO}$ is
constant.  This issue should be examined further using radial profile
data for other galaxies and using deeper CO data to determine if
molecular gas surface densities calculated using an $X_{CO}$ that
depends on metallicity are found to still be significant contributors
to the total gas surface density at larger radii and in other
galaxies.  If found to be true, then we need to more carefully
consider how to convert CO intensities into molecular gas surface
densities and possibly revise the standard depiction of the structure
of the ISM in spiral galaxies.

The bottom panel of Figure~\ref{f_radpro_gasdust} shows two radial
profiles of the gas-to-dust ratio based on the two different molecular
gas surface densities used in this analysis.  Both radial profiles
increase relatively smoothly with radius.  Because the atomic gas is
the largest component of the ISM, differing assumptions in $X_{CO}$
have only a minor impact on the radial profiles within the shown
range.  Within a kpc of the center, the ratio is seen to drop to
$\sim100$.  For comparison, the expected gas-to-dust ratio in the
Milky Way near the Earth, based either on the measurement of the
depletion of metals from the gaseous phase of the ISM or on the ratio
of gas column densities to optical extinction, is 100-200 \citep{k03,
  w03, l04, k08}.  As we indicated above, the dust masses that we have
calculated are low compared to more sophisticated models of dust
emission, so it is unlikely that we have underestimated the central
gas-to-dust ratio.  Instead, it is quite possible that the ISM in the
centre of NGC 2403 is relatively dust-rich.  The gas-to-dust ratio
reaches a value of $\sim150$ at about 2~kpc.  At radii of 5.5~kpc,
which is the limit of where we can accurately measure the radial
profile of the dust in this galaxy, the ratio reaches $\sim400$.

\begin{figure}
\begin{center}
\epsfig{file=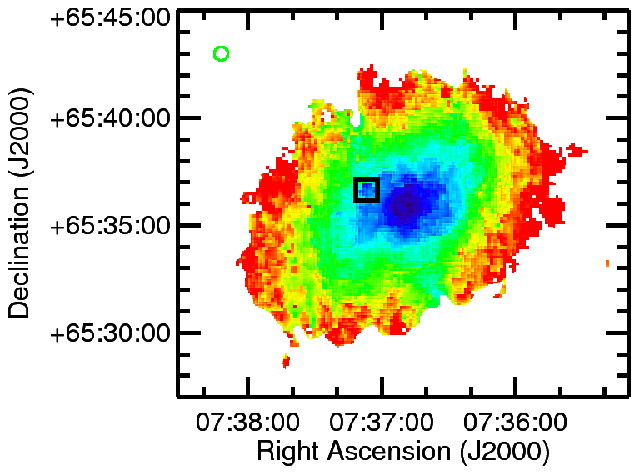}
\epsfig{file=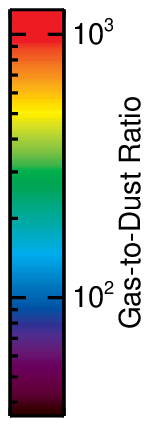}
\end{center}
\caption{Image of the gas-to-dust surface density ratio in NGC~2403.
  The image is 21~arcmin~$\times 18$~arcmin with north up and east to
  the left.  Only pixels where the dust mass was measured are shown.
  The molecular gas surface densities were calculated assuming that
  $X_{CO}$ is constant, although the functionality of $X_{CO}$ does
  not strongly affect the appearance of this figure.  The green circle
  in the upper left corner shows the 38~arcsec FWHM of the PSF, which
  is equivalent to the PSF of the 160~$\mu$m data.  The black square
  indicates the location of VS 44.}
\label{f_img_gasdust}
\end{figure}

Figure~\ref{f_img_gasdust} shows the total gas-to-dust ratio for all
pixels in which we measured dust masses.  To make this image, the data
were convolved with kernels that match the PSFs to that of the
160~$\mu$m data (which has a FWHM of 38~arcsec).  This image shows
that the gas-to-dust ratio appears to increase monotonically with
radius.  Most structures visible in Figures~\ref{f_img} and
\ref{f_img_dustmass} are not at all visible.  Even VS 44,
which has a very high surface brightness in many of the images shown
in Figure~\ref{f_img_small}, appears indistinct compared to other
locations at similar radii.  The only features that are not radially
symmetric are the streaks at the northwest and southeast edge of the
mapped data that are a result of the latent image artefacts in the
70~$\mu$m data.  The results from Figure~\ref{f_img_gasdust} suggest
that the gas-to-dust ratio is primarily dependent on radius and that
it does not vary significantly between diffuse and star-forming
regions.

Figure~\ref{f_img_gasdust} is only useful as a qualitative
demonstration of how the gas-to-dust ratio depends primarily on radius
and not on gas surface density.  As an additional quantitative
demonstration of this phenomenon, we used techniques similar to what
were used by \citet{betal08} in their comparisons of PAH 8, 24, and
160~$\mu$m emission.  We used the H{\small I}, H$_2$ (calculated using
a constant $X_{CO}$), and dust surface density maps that were used to
create Figure~\ref{f_img_gasdust}.  We rebinned these data into
45~arcsec pixels.  This pixel size was selected because it is an
integer multiple of the pixel size used for the MIPS data that is also
close to the 38~arcsec FWHM of the 160~$\mu$m data.  We then extracted
the surface densities from each 45~arcsec$^2$ region in the data.
Because the H {\small I} is the dominant component of the ISM in this
galaxy, the choice of $X_{CO}$ in this analysis has a minor impact on
the results.

\begin{figure}
\begin{center}
\epsfig{file=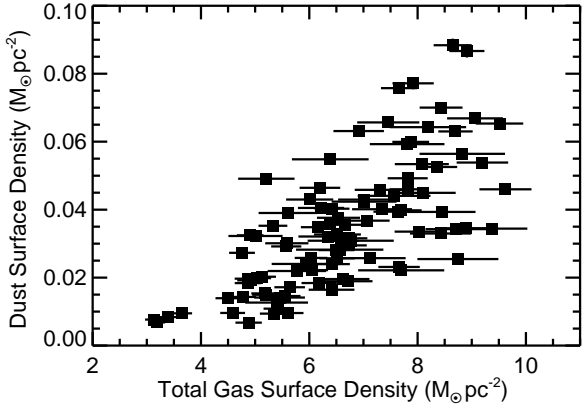}
\end{center}
\caption{The dust surface density versus the total (atomic and
  molecular) gas surface density for 45~arcsec$^2$ regions in
  NGC~2403.  The data were extracted from images that had all been
  convolved with kernels that match the PSFs to that of the
  160~$\mu$m data.  Only regions within 5.5~kpc (the limit at which
  the dust masses can be reliably measured) and with S/N
  ratios $>5$ were used.  Uncertainties in the ordinate are generally
  smaller than the symbols in this plot and the next two plots.  See
  the text in Section~\ref{s_gasdust} for a description of how these
  surface densities were extracted.}
\label{f_gasvsdust}
\end{figure}

\begin{figure}
\begin{center}
\epsfig{file=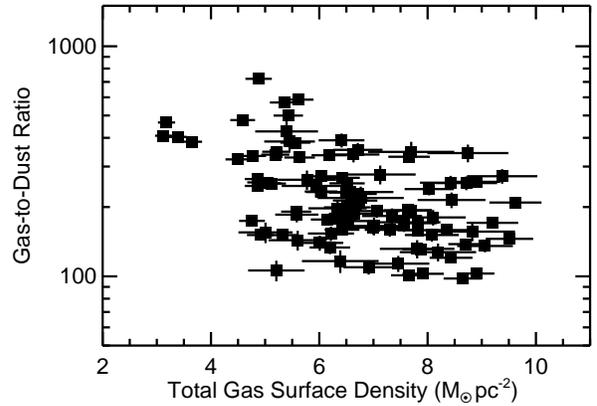}
\end{center}
\caption{The gas-to-dust ratio versus the total (atomic and molecular)
  surface density for 45~arcsec$^2$ regions in NGC~2403.  See the
  caption for Figure~\ref{f_gasvsdust} for additional information.}
\label{f_gasdustvsgas}
\end{figure}

\begin{figure}
\begin{center}
\epsfig{file=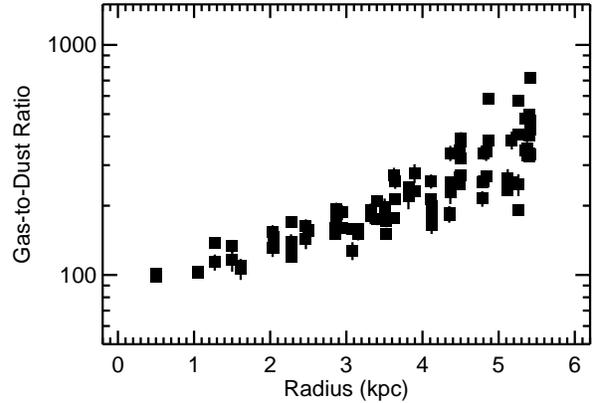}
\end{center}
\caption{The gas-to-dust ratio versus radius for 45~arcsec$^2$
  regions in NGC~2403.  This is can be considered an alternate version
  of the bottom panel in Figure~\ref{f_radpro_gasdust}.  See the
  caption for Figure~\ref{f_gasvsdust} for additional information.}
\label{f_gasdustvsrad}
\end{figure}

Figure~\ref{f_gasvsdust} shows that the total gas density is related
to the total dust density for these 45~arcsec square regions, but the
relation exhibits significant scatter.  For a given gas surface density,
the dust surface density could vary by a factor of 5.
Figure~\ref{f_gasdustvsgas} shows how the gas-to-dust ratio varies
with gas surface density.  This is an alternative form for plotting
the data in Figure~\ref{f_gasvsdust}; if a one-to-one relation existed
between the gas and dust surface densities, then the relation in
Figure~\ref{f_gasdustvsgas} should have a slope of 0 and exhibit
little scatter.  The data in Figure~\ref{f_gasdustvsgas} actually
exhibit a negative slope ($-79 \pm 13$
(M$_\odot$~pc$^{-2}$)$^{-1}$) and significant scatter.
Figure~\ref{f_gasdustvsrad} shows how the gas-to-dust ratio varies
with galactocentric radius for these 45~arcsec$^2$ regions.  The relation
here shows much less scatter; gas-to-dust ratios at a given radius may
vary by a factor of 2 or less.

Using the Spearman's rank correlation coefficient, which is a
non-parametric correlation coefficient that is useful when dealing
with data that may exhibit non-linear (i.e. exponential) relations, we
can demonstrate that the relation in Figure~\ref{f_gasdustvsrad} is
more significant than the relation in Figure~\ref{f_gasvsdust}.  For
reference, perfectly correlated data would have a coefficient of 1,
and uncorrelated data would have a coefficient of 0.  The correlation
coefficient for the relation between gas and dust surface densities in
Figure~\ref{f_gasvsdust} is 0.73, which would imply a good
correlation.  However, the correlation coefficient for the relation
between the gas-to-dust ratio and radius is 0.91.  These results
demonstrate that the relation between the gas-to-dust ratio and radius
places stronger constraints on dust surface densities than the total
gas surface density by itself.  Moreover, this quantitative approach
confirms that the absence of visible structure in
Figure~\ref{f_img_gasdust} is not simply a consequence of the poor
resolution of the data.  Instead, it is a result of the gas-to-dust
ratio being primarily dependent on radius.  However, it is always
possible that local variations in the gas-to-dust ratio will be
visible in higher-resolution data.

The most obvious driver for this gradient in the gas-to-dust ratio is
the decrease in metallicity with radius that has been observed
previously (see Section~\ref{s_data_abundance}).  The gas-to-dust
ratio is naturally expected to increase as abundances decrease simply
because fewer constituents of dust should be present in the ISM.
Gas-to-dust ratios significantly higher than the local Milky Way value
of $\sim150$ have been measured in many dwarf galaxies with low
metallicities \citep[e.g.][]{lf98, wetal07}, and globally-integrated
measurements of gas-to-dust ratios are correlated with
globally-integrated 12+log(O/H) or metallicity measurements
\citep[e.g.][]{imw90, jdee02, ddbetal07, hkos08}.  Moreover,
\citet{lf98} and \citet{metal09} have
found linear relations between oxygen abundances and dust-to-gas
ratios in nearby galaxies.  Although the dust-to-gas ratios typically
decrease faster than the oxygen abundances in these relations, the
slopes are still on the order of 1.

As stated in \ref{s_data_abundance}, the typical gradient in
12+log(O/H) that has been measured in NGC~2403 is $-0.084 \pm
0.009$~dex~kpc$^{-1}$.  For comparison, we measured the gradient in
the logarithm of the dust-to-gas ratio
($d\log(\sigma_{dust}/\sigma_{gas})/dr$) to be
$-0.097\pm0.002$~dex~kpc$^{-1}$ for NGC~2403.  These gradients are so
close that it implies that the dust-to-gas ratio is primarily
dependent on metallicity.  Alternatively, when we calculated
12+log(O/H) as a function of radius, we found the relation between
12+log(O/H) and $d\log(\sigma_{dust}/\sigma_{gas})/dr$ had a slope of
$0.86 \pm 0.02$.  This slope is less than 1, which is consistent with
the results obtained by \citet{lf98} and \citet{metal09}, but it is so
close to 1 that it implies a linear relationship between the oxygen
abundance and the dust-to-gas ratio.  Variations in the ratio of
oxygen to the other constituents of dust could explain the difference
in the gradients.  \citet{gsptsdtt99}, for example, has shown that C/O
increases systematically with O/H within this galaxy as well as within
M~101, and since carbon is a primary constituent of dust,
$d\log(\sigma_{dust}/\sigma_{gas})/dr$ would be expected to be steeper
than the gradient in 12+log(O/H).  Additional comparisons between
radially-averaged dust-to-gas ratios and 12+log(O/H) using all of the
data from SINGS, THINGS, and the JCMT NGLS would be appropriate to
determine whether similar results are typically obtained for all
spiral galaxies.
 
Some researchers \citep[e.g.][]{nggzw96, i97a} have suggested that
$X_{CO}$ for any galaxy can be determined in a two step process.
First the gas-to-dust ratio is measured in a region dominated by
atomic gas.  Second, that ratio is applied to a region that contains
CO and dust emission to determine how much molecular gas should be
present, thus giving $X_{CO}$.  However, the results here show that
the application of this method to spiral galaxies must be performed
cautiously.  The first problem is that this method assumes that the
gas-to-dust ratio is constant, whereas we have demonstrated that it
varies with radius.  The second problem is that, as explained in
Section~\ref{s_cocompare}, $X_{CO}$ may depend on metallicity, and
metallicity will vary with radius in most nearby galaxies.  Despite
these problems, it may still be possible to derive $X_{CO}$ within
subsections of galaxies by dividing the galaxies into elliptical
annuli where both the metallicity and the gas-to-dust ratio will
remain constant.  If regions free of molecular gas but containing
detectable atomic gas and dust emission can be identified within the
individual annuli, then those locations can be used to measure the
gas-to-dust ratio at specific radii.  Those regions can then be used
to infer the gas-to-dust ratio for locations at similar radii that
contain detectable CO emission to determine how much gas should be
present and therefore what $X_{CO}$ is needed to account for all of
the gas.  This will then lead to a measure of $X_{CO}$ as a function
of radius.  Unfortunately, it is not possible to apply this method to
the data used here.  To perform this analysis correctly, we would need
to work with data where the PSFs matches the 40~arcsec resolution of
the 160~$\mu$m data.  When the CO data are matched to the 160~$\mu$m
PSF, the area within radii of 4~kpc (the region in which CO~$J$=(3-2)
emission is detected along the major axis) contains almost no CO-free
regions for comparison to regions where CO is detected.

It is possible that the observations here may have missed the presence
of very cold dust (dust with temperatures $<15$~K) that may not
contribute strongly to the 70 and 160~$\mu$m bands but that may
constitute the bulk of the dust mass at larger radii.  As argued in
Appendix~\ref{a_dust}, it is unclear in cases where submillimetre
emission has been observed in excess of the $\sim20$~K thermal dust
models fit to the 70 and 160~$\mu$m data whether this excess
originates from large masses of $<15$~K dust or very low masses of
dust with enhanced submillimetre emissivities.  Results from previous
analyses have suggested that estimates of the dust mass based on the
70 and 160~$\mu$m wave bands should be well within an order of
magnitude of the expected dust mass \citep{liss02, dkw04, rtbetal04,
  betal06, ddbetal07}.  Based on these previous results, we expect the
gas-to-dust ratio shown in Figure~\ref{f_radpro_gasdust} to be
accurate and to not be strongly affected by large masses of dust not
traced by the 70 and 160~$\mu$m bands.  It is nonetheless possible
that the radial increase in our measured gas-to-dust ratio could be
linked to an increase in the relative fraction of $<15$~K dust with
radius, but to properly determine this will require both submillimetre
measurements of the dust emission and models of this dust emission
that incorporate accurate submillimetre emissivities based on
laboratory and theoretical results.

\section{Comparison of PAH, CO $J$=(3-2), and H{\small I} emission}
\label{s_pahco}

As we are interested in re-examining the empirical correlation between
CO and PAH emission found by \citet{retal06} using the
higher-sensitivity HARP-B data, and as they used radial profiles to
show this correlation, we will first look at the radial profiles for
the CO $J$=(3-2) intensity and PAH 8~$\mu$m surface brightness.  Since
\citet{retal06} used the CO intensities for their analysis instead of
converting the CO intensities to molecular gas surface densities, we
will compare the PAH 8~$\mu$m surface brightnesses directly to the CO
$J$=(3-2) intensities.

\begin{figure}
\begin{center}
\epsfig{file=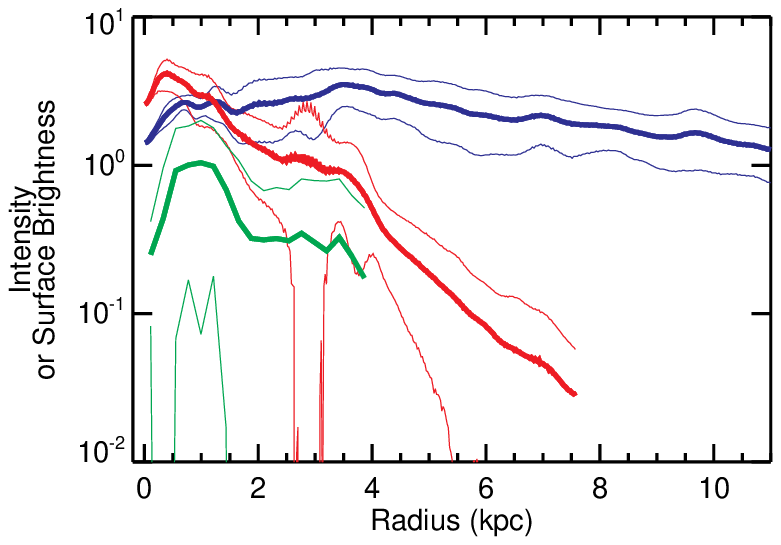}
\epsfig{file=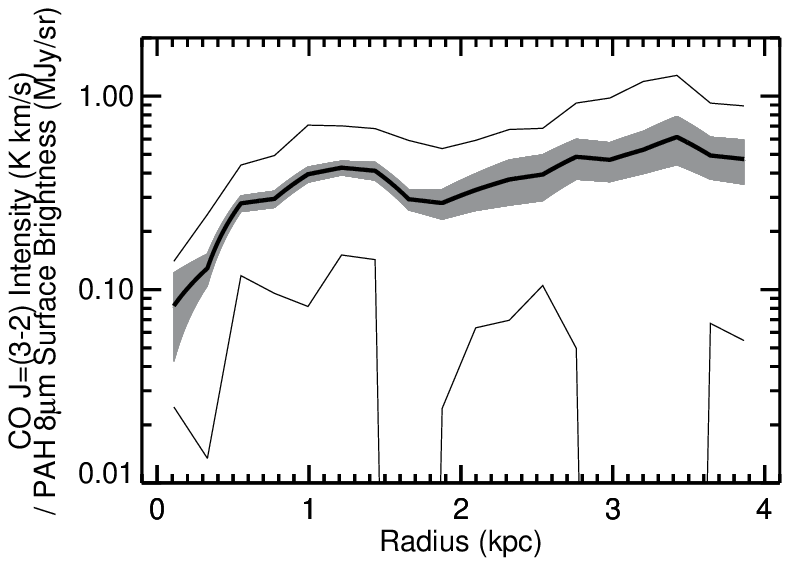}
\caption{Radial profiles of the H{\small I} intensity in Jy
  beam$^{-1}$ km s$^{-1}$ (blue line, top panel), CO $J$=(3-2)
  intensity in K km s$^{-1}$ (green line, top panel), and PAH 8~$\mu$m
  surface brightness in MJy sr$^{-1}$ (red line, top panel) as well as
  the ratio of CO $J$=(3-2) intensity to PAH 8~$\mu$m surface
  brightness (black line, bottom panel) plotted up to 10~kpc, which is
  equivalent to the edge of the optical disc.  All data were convolved
  with kernels that match the PSF of the CO $J$=(3-2) data (which has
  a FWHM of 14.5~arcsec) before radial profiles were extracted.  The
  standard deviations of the data are plotted as thinner lines around
  all radial profiles.  The uncertainties in the mean values
  are generally smaller than the thicknesses of these lines except in
  the case of the CO $J$=(3-2)/PAH 8~$\mu$m emission ratio, where the
  uncertainty is shown as a shaded grey region.  The CO $J$=(3-2) and
  PAH 8~$\mu$m radial profiles are truncated at the radii beyond which the
  quantities could be reliably measured on the major axis.  The ratio
  in the bottom panel is also truncated at the maximum radius where CO
  $J$=(3-2) intensities could be reliably measured.}
\label{f_radpro_gaspah}
\end{center}
\end{figure}

The radial profiles for the H{\small I} intensity, CO $J$=(3-2)
intensity, and PAH 8~$\mu$m surface brightness are shown in
Figure~\ref{f_radpro_gaspah}.  All data were convolved with kernels
that match the PSFs to that of the CO $J$=(3-2) data (which has a FWHM
of 14.5~arcsec) before radial profiles were extracted.  The techniques
used for extracting these radial profiles are the same as those used
in Section~\ref{s_gasdust}, although we only used data in which the CO
emission was detected to calculate the CO $J$=(3-2)/PAH 8~$\mu$m
ratio.  The width of the elliptical annuli along the major axis in the
CO~$J$=(3-2) and CO $J$=(3-2)/PAH 8~$\mu$m ratio that we used was
still equivalent to 2 pixels (or 14.6~arcsec), which does not result
in Nyquist-sampled radial profiles.  However, using narrower
elliptical profiles results in excessive noise because of the low
numbers of pixels falling within the elliptical annuli.  While the
standard deviations in the CO~$J$=(3-2) and PAH 8~$\mu$m radial
profiles are sometimes greater than the plotted radial means, just as
was the case with the CO radial profiles in
Figure~\ref{f_radpro_gasdust}, the standard deviation does not equate
to the the uncertainties in the means.  The uncertainties in the means
are actually smaller than the thickness of the lines in
Figure~\ref{f_radpro_gaspah} except for the ratio of the CO~$J$=(3-2)
intensity to PAH 8~$\mu$m surface brightness.  In effect, we can
actually measure the means to a relatively high precision,
but we still see significant variance in the data.

In the top panel of the figure, we can see that the radial profile of
the CO $J$=(3-2) and PAH 8~$\mu$m emission are similar to each other.
In contrast, the H{\small I} radial profile is much broader than
either the CO~$J$=(3-2) or PAH 8~$\mu$m radial profiles.  Given the
dissimilarity between the PAH 8~$\mu$m and H{\small I} radial profiles
as well as the dramatic differences between the appearance of the
8~$\mu$m and H{\small I} images in Figures~\ref{f_img} and
\ref{f_img_small}, we can conclude that the PAH 8~$\mu$m emission is
not associated with H{\small I} in this galaxy.  We will not discuss
the H{\small I} further but instead focus on the comparison between
the CO $J$=(3-2) and PAH 8~$\mu$m.

The CO $J$=(3-2) and PAH 8~$\mu$m radial profiles not only appear
superficially similar but also have similar exponential scale lengths.
Between 0.5~kpc and 4~kpc, the CO $J$=(3-2) radial profile has an
exponential scale length of $2.0 \pm 0.2$~kpc.  The PAH 8~$\mu$m
emission (in the convolved data) has a scale length of $1.99 \pm
0.04$~kpc over the same range.  The plot of the ratio of the two in the
bottom panel of Figure~\ref{f_radpro_gaspah} shows that the ratio
between 0.5 and 4~kpc has a gradient of $0.070 \pm 0.017$
(K~km~s$^{-1}$) (MJy~sr$^{-1}$)$^{-1}$ kpc$^{-1}$, but the mean value
for the ratio stays between $\sim0.25$ and $\sim0.6$ at radii between
0.5 and 4~kpc.  However, while the mean in the CO $J$=(3-2)/PAH
8~$\mu$m ratio remains relatively constant over this range of radii,
the high standard deviation shows that the ratio fluctuates greatly at
any given radius.  The implications of this are discussed further
below.

The CO $J$=(3-2)/PAH 8~$\mu$m emission ratio decreases sharply in the
nucleus of the galaxy.  This drop in the ratio corresponds to drops in
the surface brightness in both wave bands.  The drop in the ratio may
be in part related to variations in the gas-to-dust ratio described in
Section~\ref{s_gasdust}, but that would not explain why the ratio
appears close to constant at larger radii, nor would it explain why
the CO and PAH emission also drops near the nucleus.  We can exclude
the possibility that currently-ongoing intense nuclear star formation
may be inhibiting PAH emission (either through destroying PAHs or
through changing their ionization state) and ionizing gas in the
centre of the galaxy; both H$\alpha$ \citep[e.g.][]{drms99} and
24~$\mu$m images \citet[e.g.][]{betal08} show that the nucleus is
relatively quiescent compared to other star-forming regions which do
have PAH 8~$\mu$m and CO $J$=(3-2) counterparts.  Because this galaxy
is classified as having an H{\small II} nucleus \citep[e.g.][]{hfs97},
and no other study indicates the presence of an AGN in NGC~2403, we
can also exclude the possibility that an AGN could be responsible for
the decrease in PAH and CO $J$=(3-2) emission in the nucleus. 

The reason for the decrease in PAH 8~$\mu$m and CO $J$=(3-2) in the
nucleus of the galaxy may be explained by the simple fact that, like
many other late-type spiral galaxies, the ISM is clumpy and that this
galaxy does not contain centrally-concentrated gas and dust emission
\citep[e.g.][]{betal07}.  The clumpiness of the ISM would also explain
why the standard deviations in the PAH 8~$\mu$m and CO $J$=(3-2)
radial profiles in the top panel of Figure~\ref{f_radpro_gaspah} are
very large.  The CO $J$=(3-2)/PAH 8~$\mu$m emission ratio may decrease
towards the centre simply because the two wave bands may not be
correlated on sub-kpc scales.  The comparison of the PAH 8~$\mu$m and
CO $J$=(3-2) images in Section~\ref{s_img} visually demonstrates this;
some structures in the PAH 8~$\mu$m image were not as bright as those
in the CO $J$=(3-2) image and vice versa.  Additionally, the high
standard deviation in the radial profile of the CO $J$=(3-2)/PAH
8~$\mu$m ratio implies that the PAH 8~$\mu$m and CO emission are not
well-correlated at any given radius.

\begin{figure}
\begin{center}
\epsfig{file=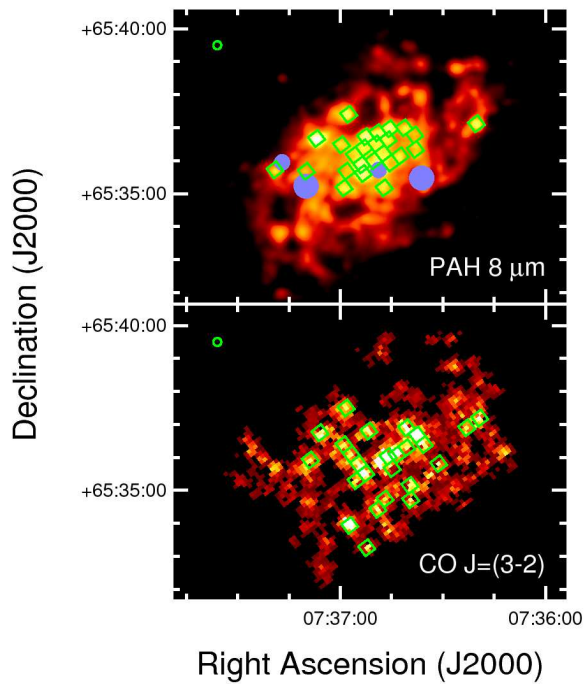}
\end{center}
\caption{Images of the PAH 8~$\mu$m and CO $J$=(3-2) emission for the
  central $12\times9$~arcmin region in NGC 2403 with green squares
  overlaid to show where intensities or surface brightnesses were
  measured.  The regions selected by their CO~$J$=(3-2) intensities
  are shown in the CO~$J$=(3-2) image, and the regions selected by
  their PAH 8~$\mu$m surface brightnesses are shown in the PAH
  8~$\mu$m image.  See the text for details on the region selection.
  The PAH 8~$\mu$m image has been convolved with a kernel to match its
  PSF to that of the CO $J$=(3-2) image.  The green circles in the
  upper left corner of each image show the 14.5~arcsec FWHM of the CO
  $J$=(3-2) PSF.  Blue regions in the PAH 8~$\mu$m image show regions
  strongly affected by stars that were excluded from the analysis in
  both wave bands.}
\label{f_pahvsco_reg}
\end{figure}

\begin{figure}
\begin{center}
\epsfig{file=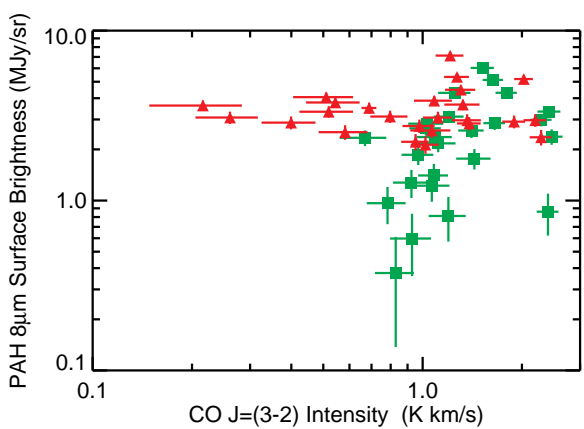}
\caption{Comparison of the CO $J$=(3-2) intensity and PAH 8~$\mu$m
  surface brightness for the regions shown in
  Figure~\ref{f_pahvsco_reg}.  The regions selected by their CO
  $J$=(3-2) intensity are shown as green squares.  The regions
  selected by their PAH 8~$\mu$m surface brightness are shown as red
  triangles.}
\label{f_pahvsco}
\end{center}
\end{figure}

For an explicit quantitative analysis to demonstrate that PAH 8~$\mu$m
and CO $J$=(3-2) emission does not trace the same structures on small
spatial scales, we could not use the same binning procedure used to
compare the dust surface density versus total gas surface density in
Section~\ref{s_gasdust}.  The gaps between detected structures in the
CO $J$=(3-2) images made the binned data difficult to interpret.  As
an alternative, we used surface brightnesses measured in a series of
discrete regions.  We selected 25 regions with the highest intensity
in the CO $J$=(3-2) image and the 25 regions with the highest surface
brightnesses in the PAH 8~$\mu$m image (convolved with a kernel to
match its PSF to the PSF of the CO $J$=(3-2) data).  These regions
were defined using the coordinate system of the CO $J$=(3-2) data;
each region is 3~pixels $\times$ 3~pixels in the CO $J$=(3-2) data
(21.8~arcsec $\times$ 21.8~arcsec or 330 pc $\times$ 330 pc).  Some of
the regions overlap, and some regions are not centered on peak-like
emission but instead selected to attempt to cover resolved structures.
The regions were also selected to avoid locations that may have been
severely affected by foreground stars.  Any locations that were
affected by foreground stars in the PAH 8~$\mu$m data were blanked out
in both data sets before surface brightnesses were extracted.  These
regions are shown in Figure~\ref{f_pahvsco_reg}.  The extracted
surface brightnesses are plotted in Figure~\ref{f_pahvsco}.

The relation between CO $J$=(3-2) and PAH 8~$\mu$m surface brightness
differs between the two sets of data, demonstrating the presence of
regions bright in one band but not the other.  The regions selected in
the PAH 8~$\mu$m data exhibit virtually no correlation; the Pearson's
correlation coefficient for these data is 0.07.  The regions selected
in the CO~$J$=(3-2) data do exhibit a weak correlation; the data have
a Pearson's correlation coefficient of 0.32.  Given that the square of
the Pearson's correlation coefficient indicates the fraction of
variation in one variable that can be accounted for by a best fit line
between the data, we find that, in the CO-selected data, only
$\sim10$\% of the variation observed in one band can be accounted for
by a relation with the other band.  In the PAH~8$\mu$m-selected data,
$<1$\% of the variation in one band can be accounted for by a relation
with the other band.  We therefore conclude that, at best, PAH
8~$\mu$m emission is only weakly correlated with CO $J$=(3-2) emission
on scales of 330~pc within this galaxy, even though the radial
profiles have very similar scale lengths.

We can reject a couple of possible scenarios that would explain why the
radial profiles for the PAH~$\mu$m and CO $J$=(3-2) emission should
match even while the emission mechanisms are not correlated on smaller
spatial scales.  First of all, PAHs and CO do not share a common
origin.  PAHs are thought to form in the atmospheres of or outflows
from AGB stars or in interstellar shocks \citep{t08}, while molecular
gas is commonly assumed to form on the surfaces of interstellar dust
grains \citep{k03, k08}.  It is also unlikely that the radial profiles
are matched through the Schmidt law, which relates star formation to
gas surface density \citep{s59, k98}.  While star formation is
strongly dependent on molecular gas surface densities in subregions
within galaxies \citep{ketal07, blwbdmt08, wetal09}, PAH emission is
not strongly linked to other star formation tracers on sub-kpc scales
\citep[e.g.][]{cetal05, cetal07, tetal07, betal08}.  Moreover,
\citet{betal08} demonstrated that the ratio of PAH 8~$\mu$m emission
to 24~$\mu$m emission (which has been shown to be correlated with
other star formation tracers) decreases as a function of radius within
NGC~2403, and the ratio of PAH 8~$\mu$m emission to other star
formation tracers would be expected to vary similarly.

On possible link between PAH and CO emission is possible if molecular
cloud formation is triggered in regions with stellar potential wells
as suggested by \citet{lwbbdmt08}.  The new molecular clouds would
produce CO emission.  Additionally, the increased starlight in these
potential wells (from both the stars originally present in the wells
and new stars formed from the molecular clouds) would heat the diffuse
ISM.  According to the results from \citet{betal08}, enhanced PAH
emission will be produced by locations where the interstellar
radiation field has increased.  Therefore, the stellar potential wells
that form molecular gas clouds in this scenario would not only enhance
CO emission but also enhance PAH emission, thus linking the two on
large spatial scales and resulting in radial profiles that appear
similar for both wave bands.  However, the enhanced PAH emission does
not need to occur within the stellar potential wells but could instead
occur in regions surrounding the potential wells.  Therefore, regions
containing molecular gas could exhibit CO emission correlated with PAH
emission while regions without molecular gas would exhibit PAH
emission that is uncorrelated with CO emission as shown in
Figure~\ref{f_pahvsco}.

It is also possible that both PAH and CO emission are affected by
similar excitation mechanisms.  As stated above, enhancements in PAH
emission may be indicative of increased heating of the diffuse ISM,
which could be caused by enhancing star formation activity (although
they still declared that PAH emission was not useful for tracing star
formation on sub-kpc scales).  Similarly, \citet{kb96} postulated
based on the comparison of CO radial profiles to other star formation
tracers that CO could be excited by the cosmic rays from star
formation, and cosmic rays tend to diffuse into the ISM as can be seen
by the comparisons of infrared and radio continuum emission performed
by \citet{metal06a}, \citet{metal06b}, and \citet{mhkab08}.
Therefore, increases in star formation could enhance both PAH and CO
emission on large spatial scales, which would explain why the radial
profiles for PAH and CO emission appear to match each other.

It is also possible that the radial profiles of CO~$J$=(3-2) and PAH
8~$\mu$m emission are similar simply by coincidence.  The radial
profiles of both normalized by the total gas surface density are
expected to decrease with radius, although for different reasons.
Perhaps the multiplicative factors that relate the slope of the
CO~$J$=(3-2) radial profile to that of total gas surface density just
happen to match those factors that relate the PAH 8~$\mu$m radial
profile to that for total gas surface density, at least in this galaxy
and the ones studied by \citet{retal06}.  A more extensive
investigation using the whole of the JCMT NGLS sample could
potentially reveal galaxies where the PAH 8~$\mu$m and CO~$J$=(3-2)
radial profiles are not similar (i.e. where the radial profiles do not
have similar exponential scale lengths).  If a sufficient number of
such galaxies can be identified, then it would indicate that what was
found here was merely a coincidence.

\section{Conclusions}

We find that the dust surface density depends on the total (atomic and
molecular) gas surface density and on radius.  The gas-to-dust ratio
varies from $\sim100$ in the nucleus to $\sim400$ at 5.5~kpc, which is
the limit at which the dust surface densities can be detected.  The
logarithm of the slope for the gas-to-dust ratio
($d\log(\sigma_{dust}/\sigma_{gas})/dr$)), which is
$0.097\pm0.002$~dex~kpc$^{-1}$, is very close to the slope in
12+log(O/H), which is usually reported as between -0.07 to
-0.09~dex~kpc$^{-1}$.  This implies that the gas-to-dust ratio is
strongly dependent on metallicity.  Additional abundance variations,
such as variations in the C/O ratio, could explain the small disparity
between $d\log(\sigma_{dust}/\sigma_{gas})/dr$ and 12+log(O/H).  Aside
from these strong radial variations in the gas-to-dust ratio, the
ratio does not appear to vary within the galaxy.

We also find that the PAH 8~$\mu$m and CO~$J$=(3-2) radial profiles
have statistically identical scale lengths.  On spatial scales of
220~pc, however, PAH 8~$\mu$m and CO~$J$=(3-2) emission are
uncorrelated.  We propose two mechanisms that could link the PAH
emission with star formation.  First, the CO emission may appear
associated with PAH emission if molecular cloud formation is triggered
by stellar potential wells.  The stars in these potential wells would
enhance the radiation field that heats the diffuse ISM, which would
enhance PAH emission.  Thus, the PAH emission would appear associated
with CO emission on large scales but not on small scales.  Second, the
two wave bands may be related through different excitation mechanisms
that are linked through star formation and that affect the ISM on
large areas.  While PAH emission does not trace star formation on
sub-kpc scales, it is expected to be enhanced by increases in the
interstellar radiation field that accompany enhanced star formation.
Meanwhile, CO emission could be enhanced by the increase in cosmic
rays that accompanies enhanced star formation.  Thus,
radially-averaged PAH and CO emission would both be linked through
star formation, even though PAH and CO emission may not be spatially
correlated on sub-kpc scales.  However, we also do not rule out the
possibility that the may just coincidentally have radial profiles
with similar scale lengths.

This paper is only a first look at the gas-to-dust ratio and the
relation between PAH and CO emission within the JCMT NGLS sample.  The
SINGS, THINGS, and JCMT NGLS samples were all selected to contain many
of the same galaxies.  The subset of galaxies found in all three
samples can be used for an extended comparison of dust, molecular gas,
and atomic gas surface densities and for comparisons of PAH to CO
emission.  Moreover, the gradients in the gas-to-dust ratio could then
be compared to abundance gradients to examine the relation
between the two further.

\section*{Acknowledgments}

We thank the reviewer for his/her comments on this paper.  The James
Clerk Maxwell Telescope is operated by The Joint Astronomy Centre on
behalf of the Science and Technology Facilities Council of the United
Kingdom, the Netherlands Organisation for Scientific Research, and the
National Research Council of Canada.  G.J.B. and D.L.C. were funded by
STFC.  The research of J.I.  and C.D.W. is supported by grants from
NSERC (Canada). A.U. has been supported through a Post Doctoral
Research Assistantship from the UK Science \& Technology Facilities
Council.  Travel support for B.E.W. and T.W. was supplied by the
National Research Council (Canada).  This research has made use of the
NASA/IPAC Extragalactic Database (NED) which is operated by the Jet
Propulsion Laboratory, California Institute of Technology, under
contract with the National Aeronautics and Space Administration.

\appendix
\section{Discussion on dust not traced by the MIPS data}
\label{a_dust}

It is entirely possible that the 70 and 160~$\mu$m bands may have
missed significant masses of very cold dust (dust with temperatures
$<15$~K) within this galaxy.  However, it is unclear whether such dust
is present.  

Many surveys working with {\it Spitzer} and submillimetre data have
not found evidence for very cold dust.  The best examples are the
results from \citet{detal05} and \citet{ddbetal07}, who worked with
{\it Spitzer} and submillimetre data for 17 spiral and S0 galaxies in
the SINGS sample and found no submillimetre emission in excess of
what was predicted by dust models that did not include $<15$~K dust.
Unfortunately, the quality of the submillimetre data used in these
studies were quite variable, so it may not have been feasible for
\citet{detal05} and \citet{ddbetal07} to detect submillimetre emission
in excess of what is expected from $>15$~K dust within most galaxies.
A few groups using very high signal-to-noise submillimetre data with
ISO or {\it Spitzer} data have conclusively demonstrated the presence
of emission at wavelengths $>850$~$\mu$m in excess of what is expected
from $\sim20$~K dust within the Milky Way Galaxy \citep{retal95,
  fds99}, NGC~4631 \citep{dkw04, betal06}, NGC~3310 \citep{zpxkl09},
and a few nearby dwarf galaxies \citep{liss02, gmjwbl03, gmjwb05}.
While \citet{gmjwbl03} and \citet{gmjwb05} have suggested that this
represents emission from a high mass of dust at 5-10~K, most other
authors have rejected this explanation because it leads to implausibly
low gas-to-dust ratios.  Alternatively, the submillimetre emissivity
of dust may be enhanced above the $\lambda^{-2}$ emissivity law
commonly used in many dust models \citep[e.g.][]{ld01}.  Such dust
would include grains with exotic shapes, such as fractal dust grains,
which would have relatively high submillimetre emissivities relative
to their absorption cross sections \citep{retal95, dkw04, betal06};
dust in which resonances from impurities in the dust grains enhances
the submillimetre emissivity \citep{retal95, betal06}; or grains with
shallow dust emissivities ($\kappa_{\nu} \propto \lambda^{-\beta}$,
with $\beta < 2$) that would only be visible at submillimetre
wavelengths \citep{fds99, liss02, zpxkl09}.  Additionally, laboratory
studies \citep{asjbb96, mbcprb98, bmnjbhm05} and at least one
theoretical study \citep{mgbbpn07} have shown that dust grains exhibit
a broad range of emissivities at these wavelength and that
submillimetre dust emissivities can be strongly enhanced above the
$\lambda^{-2}$ emissivity function that is commonly used in
astronomical research.  In all of these alternate scenarios, the mass
of the dust that produces the excess submillimetre emission is
relatively low, and most of the dust mass should be traced by dust
that dominates the SED at $<450$~$\mu$m.  Moreover, the results from
\citet{liss02}, \citet{dkw04}, \citet{rtbetal04}, \citet{betal06}, and
\citet{ddbetal07} have demonstrated that, based on the total gas
content and expected gas-to-dust ratios for nearby galaxies, most of
the dust mass in nearby galaxies can be accounted for by $\gtrsim
15$~K dust emitting at $\geq 450$~$\mu$m.

It is always possible that newer $>160$~$\mu$m data from either the
{\it Herschel} Space Observatory or SCUBA2 on the JCMT along with
improved dust models could demonstrate that the use of {\it Spitzer}
data alone will lead to significant underestimates of dust masses.
However, based on current data and models, the dust masses that we
measure using only the {\it Spitzer} data should provide a reasonably
accurate estimate of the dust masses.  We should not be
severely underestimating the dust masses because we have not accounted
for large masses of cold dust emitting $>160$~$\mu$m.

\label{lastpage}

\end{document}